# Finite-sample adjustments for comparing clustered adaptive interventions using data from a clustered SMART

Wenchu Pan[1,*], Daniel Almirall[2,3], Amy M. Kilbourne[4], Andrew Quanbeck[5], and Lu Wang[1]

[1]Department of Biostatistics, University of Michigan, Ann Arbor, MI, USA

[2]Survey Research Center, University of Michigan, Ann Arbor, MI, USA

[3]Department of Statistics, University of Michigan, Ann Arbor, MI, USA

[4]Department of Learning Health Sciences, University of Michigan Medical School, Ann Arbor, MI, USA

[5]Department of Family Medicine and Community Health, University of Wisconsin-Madison, WI, USA

*email: wenchu@umich.edu,    corresponding author

SUMMARY:    Adaptive interventions, aka dynamic treatment regimens, are sequences of pre-specified decision rules that guide the provision of treatment for an individual given information about their baseline and evolving needs, including in response to prior intervention. Clustered adaptive interventions (cAIs) extend this idea by guiding the provision of intervention at the level of clusters (e.g., clinics), but with the goal of improving outcomes at the level of individuals within the cluster (e.g., clinicians or patients within clinics). A clustered, sequential multiple-assignment randomized trials (cSMARTs) is a multistage, multilevel randomized trial design used to construct high-quality cAIs. In a cSMART, clusters are randomized at multiple intervention decision points; at each decision point, the randomization probability can depend on response to prior data. A challenge in cluster-randomized trials, including cSMARTs, is the deleterious effect of small samples of clusters on statistical inference, particularly via estimation of standard errors.

This manuscript develops finite-sample adjustment (FSA) methods for making improved statistical inference about the causal effects of cAIs in a cSMART. The paper develops FSA methods that (i) scale variance estimators using a degree-of-freedom adjustment, (ii) reference a t distribution (instead of a normal), and (iii) employ a "bias corrected" variance estimator. Method (iii) requires extensions that are unique to the analysis of cSMARTs. Extensive simulation experiments are used to test the performance of the methods. The methods are illustrated using the Adaptive School-based Implementation of CBT (ASIC) study, a cSMART designed to construct a cAI for improving the delivery of cognitive behavioral therapy (CBT) by school mental health professionals within high schools in Michigan.

This paper has been submitted for consideration for publication in *Biometrics*

## 1. Introduction

Adaptive interventions (AI), as applied to individuals (e.g., patients), are sequences of pre-specified decision rules that guide the provision of treatment given information about the baseline information and evolving needs of individuals (1; 4; 5). Adaptive interventions are also known as dynamic treatment regimens (6; 7).

Adaptive interventions can also guide the provision of intervention for clusters of individuals (e.g., clinics comprised of patients, or schools comprised of teachers). In this case, we call them clustered adaptive interventions (cAIs) (9; 31; 32; 33). In cAIs, the sequence of decision rules guide the provision of intervention given the baseline and ongoing needs of each cluster. Cluster-level intervention is tailored based on cluster-level factors (e.g., school or clinic factors), including cluster-level summaries (aggregates) of individual-level factors (e.g., measures of expertise of clinicians at a clinic). As with adaptive interventions, a common goal of a cAI is to improve outcomes at the level of the individuals within clusters (e.g., teachers within the school).

Clustered, sequential multiple-assignment randomized trials (cSMARTs) (1; 2; 3; 8; 9) are one type of multistage randomized trial that can be used to construct high-quality cAIs. In a cSMART, clusters are randomized at multiple critical decision points; at each decision point, the randomization probability can depend on response to prior intervention. The defining feature of a cSMART is that the sequence of randomizations is at the cluster-level (e.g. school or clinic), whereas the trial's primary outcome is at the level of units within the cluster (e.g. clinicians within the clinic or school). cSMARTs have been applied across a wide range of fields, including mental health (11; 12), opioid use disorder prevention (10), education and others (13; 14; 15).

A common primary aim in a cSMART concerns marginal mean comparison between two (or more) cAIs (9; 11). Building on work by Chakraborty, et al. (7), Orellana et al (17), and Nahum-Shani et al (5), Necamp, et al (9) developed a marginal mean regression approach with weighted least squares estimation for comparing cAIs in a cSMART.

However, a common problem in any clustered randomized trial, including cSMARTs, is the deleterious effect of small numbers of clusters. When the number of clusters is small, these existing methods can lead to underestimated variance, smaller-than-nominal confidence intervals, and hypothesis tests with greater-than-nominal type I errors.

The primary contribution of this manuscript is the development of finite-sample adjustment (FSA) methods for making improved statistical inference about the relative causal effect between cAIs using data from a cSMART. We develop and test methods for cSMARTs that (i) scale variance estimators using a degree-of-freedom adjustment, (ii) reference a $t$ distribution (instead of a normal), (iii) employ a "bias-corrected" variance estimator, and the combinations of these methods. Methods (i) and (ii) can be extended for cSMARTs in a straightforward manner without the use of special software. The bias-corrected method (iii), on the other hand, requires an extension (and specialized software) to accommodate the fact that some clusters in a cSMART contribute data for multiple cAIs. Extensive simulation experiments are used to understand the operating characteristics of the three approaches, justify the performance improvements. To our knowledge, this is the first manuscript to develop and test the performance of FSA using data from cSMARTs.

SMARTs are multi-stage randomized trial designs used to inform the construction of high-quality adaptive interventions (1; 2; 3; 8; 9). This manuscript is motivated by the Adaptive School-based Implementation of CBT (ASIC) study (11), described below. All methods presented here can be extended easily for use with data from other types of cSMARTs, including with more than two stages of randomization or unequal randomization probabilities.

## 2. Clustered Sequential Multiple-Assignment Randomized Trials

2.1 *The ASIC Study*

Figure 1 describes ASIC (11), which is an example of a prototypical, two-stage cSMART design, whereby: (i) all clusters (schools) are randomized with equal probability to two first-stage intervention options (Coaching vs No Coaching); (ii) all clusters are monitored for their response to the first-stage intervention; (iii) non-responding clusters are re-randomized with equal probability to two second-stage intervention options (Adding Facilitation vs No Facilitation); and (iv) responding clusters are not re-randomized (e.g., continue with first-stage intervention). In ASIC, a school is categorized as "non-responding" if: (i) at least one school professional (SP) within the school delivers $< 3$ cognitive behavioral therapy (CBT) components to $< 10$ students, or (ii) school professional report, on average, $> 2$ barriers to CBT uptake.

2.2 *Notation*

2.2.1 *Observed Data.* Let $i$ denote a cluster, where $i = 1, \ldots, n$. $X_i$ denotes a $p$-dimensional baseline (pre-randomization) vector of cluster-level covariates. $A_{1i} \in (-1, 1)$ denotes the assigned first-stage intervention option. $R_i \in (0, 1)$ denotes response status following $A_{1i}$. For $R_i = 0$, $A_{2i} \in (-1, 1)$ denotes the assigned second-stage intervention option. For $R_i = 1$, $A_{2i}$ is missing by design; i.e., $A_{2i} =$ NA.

In a prototypical cSMART there are $q = 4$ embedded cAIs (i.e., 2 randomizations $\times$ 2 intervention options at each randomization); see Table 1. Each cAI can be denoted $(a_1, a_{2R}, a_{2NR})$, where $a_1$ denotes first-stage intervention, $a_{2R}$ denotes second-stage intervention for responders, and $a_{2NR}$ denotes second-stage intervention for non-responders. Given $a_1$, however, there is no variation in $a_{2R}$. To avoid unnecessary notation, we omit $a_{2R}$; and use the simpler notation $(a_1, a_{2NR}) \equiv (a_1, a_2) \in \mathcal{A}$ where $\mathcal{A} = \{(1, 1), (1, -1), (-1, 1), (-1, -1)\}$.

For example, in ASIC, cAI (-1,-1) is the intervention that provides stage 1 Coaching to all schools, continues with Coaching for responding schools, and augments with Facilitation for non-responding schools.

The random variable $Y_{ij}$ denotes the primary research outcome (continuous) for participant (school professional) $j$ nested within cluster (school) $i$ ($j = 1, \ldots, m_i$); and the vector $Y_i = \{Y_{i1}, \ldots, Y_{im_i}\}$ denotes the collection of observed outcomes within cluster $i$. $Y_i$ is measured at a common endpoint following second-stage assignment.

2.2.2 *Potential Outcomes.*   Potential outcomes notation is used to define the primary aim (i.e., causal estimand) in a cSMART ((26)). $R_i(a_1)$ denotes the end of stage 1 response status had cluster $i$ been assigned stage 1 $a_1$. $Y_i(a_1, a_2)$ denotes the vector of potential outcomes had cluster $i$ been assigned AI$(a_1, a_2)$. For each cluster $i$ there are 2 potential response status outcomes $R_i(a_1)$; and 4 potential primary outcomes $Y_i(a_1, a_2)$.

2.2.3 *Complete Data.*   The complete observed data is $\{X_i, A_{1i}, R_i, A_{2i}, Y_i : i = 1, \ldots, n\}$. The complete potential outcomes are $\{R_i(a_1), Y_i(a_1, a_2) : (a_1, a_2) \in \mathcal{A}, \ i = 1, \ldots, n\}$. The total sample size is $N = \sum_i^n m_i$ where $n$ is the total number of clusters. When there is no potential for confusion, for brevity we drop subscript $i$ for the observed or potential outcomes.

2.3 *Primary Aims (Causal Estimands)*

$E(Y(a_1, a_2))$ denotes the marginal mean of $Y$ had the entire population been assigned $(a_1, a_2)$. Typical primary aims in a cSMART concern pairwise comparisons between cAIs (27; 28):

$$\Delta_{\begin{bmatrix}(a_1,a_2)\\(a_1',a_2')\end{bmatrix}} = E(Y(a_1, a_2) - Y(a_1', a_2')), \text{ for any fixed pair of cAIs } (a_1, a_2) \neq (a_1', a_2'). \quad (1)$$

For example, in ASIC, the primary aim was a test of the null that $H_0 : \Delta_{\begin{bmatrix}(1,1)\\(-1,-1)\end{bmatrix}} = 0$; i.e., no average causal effect on $Y$ (the average number of CBT sessions delivered by week 33) between REP+COACH+FAC $(-1, -1)$ versus REP only $(1, 1)$.

2.4 *Marginal Mean Modeling*

To facilitate a regression approach (16), we consider pre-specified marginal mean models for $E(Y(a_1, a_2) \mid X)$ of the form

$$\mu(a_1, a_2, X; \beta, \eta) = \beta_0 + \beta_1 a_1 + \beta_2 a_2 + \beta_3 a_1 a_2 + \eta^T \bar{X}, \qquad (2)$$

where $\bar{X}$ is the vector of mean-centered covariates $X$—note $E(\bar{X}) = 0$—and $(\beta, \eta)$ is a $q + p$ vector of unknown parameters. Baseline covariates are often included because they may lead to improved statistical efficiency in the comparison of cAIs ((46; 47)). (Note: $\bar{X}$-by-$(a_1, a_2)$ interaction terms are allowed, but omitted here for simplicity.) For example, based on (2), the primary aim in ASIC corresponds to testing the null that $E_X(E(\mu(1, 1, X; \beta, \eta) - \mu(-1, -1, X; \beta, \eta))) = 2\beta_1 + 2\beta_2$ is equal to zero.

All parameters have scientific interpretation: $\beta_0$ is the grand mean of $Y$, marginal over $(X, A_1, R, A_2)$. $2\beta_1$ represents the main causal effect of cAIs that assign first-stage $a_1 = 1$ vs $a_1 = -1$, marginal over $(X, R, A_2)$. $2\beta_2$ is the main causal effect of cAIs that assign second-stage $a_2 = 1$ vs $a_2 = -1$, marginal over $(X, A_1, R)$. Non-zero $\beta_3$ indicates the presence of an interaction effect between $a_1$ and $a_2$; i.e., the joint effect of $a_1$ and $a_2$ is greater/lesser than the sum of its main effects. $\eta$ is the (non-causal) association between $X$ and $Y$; e.g., $\eta \times \text{sd}(X)/\text{sd}(Y)$ approximates the correlation of $X$ and $Y$ marginal over $(A_1, R, A_2)$.

## 3. Methodology

3.1 *Estimation and Inference*

Let the "sequence" $S_i = \{A_{1i}, R_i, A_{2i}\}$ denote the triplet identifying cluster $i$'s observed intervention-response pathway. Following (1; 7; 29), we estimate $(\beta, \eta)$ as the solution to $0 = \mathbb{P}_n U_{\beta,\eta}(X, S, Y; \beta, \eta)$, where $\mathbb{P}_n(Z)$ is shorthand for $\frac{1}{n}\sum_i^n Z_i$ for any $Z$, and the $(q+p)$- dimensional vector $U_{i,\beta,\eta}$ is given by

$$U_{i,\beta,\eta} = \sum_{(a_1,a_2)} I_{i,a_1,a_2}(S_i) W_{i,a_1,a_2}(X_i, S_i) D_{a_1,a_2}^T(X_i) V_{i,a_1,a_2}^{-1}(X_i)(Y_i - \mu(a_1, a_2, X_i; \beta, \eta)). \quad (3)$$

The indicator $I_{i,a_1,a_2}(S_i)$ is used to denote whether ($= 1$) or not ($= 0$) cluster $i$'s observed intervention-response sequence ($S_i$) is consistent with cAI ($a_1, a_2$). When combined with the sum over the cAIs, $I_{i,a_1,a_2}(S_i)$ is used to select the data used to estimate each $E(Y(a_1, a_2))$.

For example, in a prototypical cSMART: For responders to $A_{1i} = a_1$, the sequence $S_i$ is given by $(A_{1i} = a_1, R_i = 1, A_{2i} = \text{NA})$; so the $(X_i, Y_i)$ data from responders to $a_1$ is used to estimate the mean under $(a_1, 1)$ and $(a_1, -1)$. On the other hand, for non-responders to $A_{1i} = a_1$, the sequence $S_i$ is given by $(A_{1i} = a_1, R_i = 0, A_{2i} = a_2)$; so the $(X_i, , Y_i)$ data from non-responders to $a_1$ is used to estimate the mean under $(a_1, a_2)$.

The weights $W_{i,a_1,a_2}(X_i, S_i)$ equal the inverse of the product of the known $p_{1i,a_1}(X_i) = Pr(A_{1i} = a_1 \mid X_i)$ and $p_{2i,a_1,a_2}(X_i, R_i) = Pr(A_{2i} = a_2 \mid X_i, A_{1i} = a_1, R_i)$. In the prototypical cSMART, $p_{1i,a_1}(X_i) = p_{2i,a_1,a_2}(X_i, R_i = 0) = 1/2$; here, $W_{i,a_1,a_2}(X_i, S_i) = 2R_i + 4(1 - R_i)$.

$D_{a_1,a_2}(X_i)$ is the $m_i \times (q+p)$ design matrix representing the derivative of $\mu(a_1, a_2, X_i; \beta, \eta)$ with respect to $(\beta, \eta)$; e.g., using (2), the $j_{th}$ row of $D_{a_1,a_2}(X_i)$ is equal to $(1, a_1, a_2, a_1 a_2, \bar{X}_j^T)$.

$V_{i,a_1,a_2}(X_i)$ is an $m_i \times m_i$ matrix for the variance of the residual vector $\epsilon_{i,(a_1,a_2)}(X_i, Y_i; \beta, \eta)$ given $X_i$, where $\epsilon_{i,(a_1,a_2)}(X_i, Y_i; \beta, \eta) = (Y_i(a_1, a_2) - \mu(a_1, a_2, X_i; \beta, \eta))$.

Appendix 2 (and (20)) shows that, as $n \to \infty$, $(\hat{\beta}, \hat{\eta})$ is normally distributed with mean $(\beta, \eta)$ (unbiased) and asymptotic variance given by

$$\text{var}(\hat{\beta}, \hat{\eta}) = \Sigma_{\beta,\eta} = \frac{1}{n} B^{-1} M B^{-1}, \text{ a } (p+q) \times (p+q) \text{ matrix, where} \quad (4)$$

$$M = E(U_{i,\beta,\eta} U_{i,\beta,\eta}^T), \quad B = E(\sum_{(a_1,a_2)} I_{i,a_1,a_2}(S_i) W_{i,a_1,a_2}(S_i) D_{a_1,a_2}(X_i)^T V_{i,a_1,a_2}^{-1} D_{a_1,a_2}(X_i)). \quad (5)$$

These results hold for any (invertible) working model for $V_{i,a_1,a_2}$. The upper-left $q \times q$ submatrix in $\Sigma_{\beta,\eta}$ forms the basis for standard errors, confidence intervals, and hypothesis tests for the causal $\beta$. In Section 4 we introduce finite sample adjustments to $\Sigma_{\beta,\eta}$.

3.2 *Implementation*

3.2.1 *Working Models for $V_{i,a_1,a_2}(X_i)$.* Non-independent working models for $V_{i,a_1,a_2}$ are often pre-specified because they may lead to improved statistical efficiency. Let $\alpha$ be a finite vector of unknown parameters indexing a working model for the variance, now denoted $V_{i,a_1,a_2}(X_i, \alpha)$. For example, a common choice is the compound symmetric structure $V_{i,a_1,a_2}(X_i; \alpha) = \sigma^2 CS(\rho)$, where $CS(\rho)$ is an $m_i \times m_i$ matrix with 1 on the diagonal and $\rho$ in off-diagonals; here, $\alpha = (\sigma^2, \rho)$; and $\rho$ is the intra-cluster correlation (ICC; (48)). $V_{i,a_1,a_2}(X_i; \alpha)$ may depend on $X_i$, employ heterogeneous $\sigma^2_{a_1,a_2}$ or $\rho_{a_1,a_2}$, or use a different structure altogether.

$V_{i,a_1,a_2}(X_i; \alpha)$ is a working model for $\text{var}(\epsilon_{i,(a_1,a_2)}(X, Y; \beta, \eta) \mid X)$ which depends on $(\beta, \eta)$. Pseudo-algorithm 1 describes an approach to alternating between $(\widehat{\beta}, \widehat{\eta})$ and $\widehat{\alpha}$ (shown for the example working model $V_{i,a_1,a_2}(X_i; \alpha) = \sigma^2 CS(\rho)$). This approach is similar to a constrained pseudo-maximum likelihood estimator (see Appendix 1) and extends easily to other working models.

---
**Pseudo-algorithm 1** Iterative Procedure for the Compound Symmetry Example

**initialize**
  Set $\iota = 0$; $\widehat{\sigma}^2_{a_1,a_2}[\iota] = 1$; and $\widehat{\rho}_{a_1,a_2}[\iota] = 0$
**repeat** $\iota = 0, \ldots$
  **Step 1** Obtain $\hat{\beta}[\iota], \hat{\eta}[\iota]$ by solving Equation (3)
  **Step 2** For each $AI(a_1, a_2)$, calculate $\widehat{\epsilon}_{ij,(a_1,a_2)} = Y_{ij} - \mu(X_{ij}, a_1, a_2; \hat{\beta}[\iota], \hat{\eta}[\iota])$
  **Step 3** Estimate $\widehat{\sigma}^2_{a_1,a_2}[\iota]$ and $\widehat{\rho}_{a_1,a_2}[\iota]$ using
  $$\widehat{\sigma}^2_{a_1,a_2}[\iota] = \frac{\sum_{i=1}^{n}[W_i I_{i,a_1,a_2} \sum_{j=1}^{m_i} \hat{\epsilon}^2_{ij,(a_1,a_2)}]}{\sum_{i=1}^{n} W_i I_{i,a_1,a_2} m_i}, \quad \text{and}$$
  $$\widehat{\rho}_{a_1,a_2}[\iota] = max\left\{\frac{\sum_{i=1}^{n}[W_i I_{i,a_1,a_2} \sum_{j=1}^{m_i}\sum_{k \neq j}^{m_i} \hat{\epsilon}_{ij,(a_1,a_2)}\hat{\epsilon}_{ik,(a_1,a_2)}]}{\hat{\sigma}^{2*}_{a_1,a_2} \sum_{i=1}^{n} W_i I_{i,a_1,a_2} m_i(m_i-1)}, 0\right\}$$
  **Step 4** $\iota \leftarrow \iota + 1$
**until** $(\hat{\beta}[\iota], \hat{\eta}[\iota])$ converges

---

3.2.2 *Estimated Weights.* $W_{i,a_1,a_2}(X_i, S_i)$ is known function of the randomization probabilities. Yet, estimating the weights can lead to improved statistical efficiency (34; 35). Let $\gamma$ be a finite vector of parameters indexing a model for the probabilities, i.e., $(p_{1i,a_1;\gamma_1}, p_{2i,a_1,a_2;\gamma_2})$, with associated model-based weights denoted by $W_{i,a_1,a_2}(X_i, S_i; \gamma)$. Let $\hat{\gamma}$ be any regular estimator for $\gamma$. For example, in the prototypical cSMART, using an intercept-only logistic regression for $A_1$ and $A_2 \mid (A_1, R = 0)$ leads to $\widehat{p}_{1i,a_1} = \mathbb{P}_n I(A_{1i} = a_1)$ and $\widehat{p}_{2i,a_2} = \mathbb{P}_n I(A_{2i} = a_2)I(R_i = 0)/\mathbb{P}_n I(R_i = 0)$. Models may also include baseline or time-varying covariates. Let $S_{\gamma,i}$ denote the score vector of the estimating equation for solving $\hat{\gamma}$. To account for the estimation of $\gamma$ in the standard errors, we use:

$$\mathrm{var}(\hat{\beta}, \hat{\eta}) = \Sigma_{\beta,\eta} = \frac{1}{n} B^{-1}(M - CF^{-1}C)B^{-1}, \text{ where} \qquad (6)$$

$$C = E\left(\frac{\partial U_{i,\beta,\eta}}{\partial \gamma}\right), \text{ and } F = E(S_{\gamma,i} S_{\gamma,i}^T). \qquad (7)$$

See Appendix 5 for a proof.

3.2.3 *Estimated Standard Errors.* A plug-in estimator is used to obtain the estimated var-cov matrix $\widehat{\Sigma}_{\beta,\eta}$ by (i) replacing all expectations $E$ in displays (4)-(7) with the empirical average $\mathbb{P}_n$; and (ii) plugging in $(\beta, \eta, \alpha, \gamma)$ for $(\hat{\beta}, \hat{\eta}, \hat{\alpha}, \hat{\gamma})$.

## 4. Finite Sample Adjustments

The justifications for $\widehat{\Sigma}_{\beta,\eta}$ and use of a Normal reference distribution is asymptotic. However, the literature on standard cluster-randomized trials ((38)) shows that, in small samples, such variance estimators are biased downward, and confidence intervals using $\widehat{\Sigma}_{\beta,\eta}$ and the Normal distribution can have substantially lower-than-nominal coverage rates.

Below, we describe four finite sample adjustments (FSA): FSA1 is used in all situations henceforth because, without it, non-independent $\widehat{V}_{a_1,a_2}$ cannot be inverted in cSMARTs with small samples; this is the only FSA related to the estimators $(\widehat{\beta}, \widehat{\eta})$. FSA2 concerns the reference distribution used to make inference; whereas, FSA's 3 and 4 involve direct adjustments to $\mathrm{var}(\hat{\beta}, \hat{\eta}) = \Sigma_{\beta,\eta}$. Whereas FSA's 1-3 are relatively straightforward to implement (e.g., with data from any type of randomized trial), FSA4 required extensions to $\Sigma_{\beta,\eta}$ to accommodate the unique features of a cSMART.

### 4.1 Enforcing Nonnegative Correlation (FSA1)

FSA1 forces $\widehat{\rho}_{a_1,a_2}$ to be non-negative (as shown in Pseudo-algorithm 1). In the applications envisioned, $\rho_{a_1,a_2}$ is expected to be non-negative ((36)). Further, in simulations where $\widehat{\rho}_{a_1,a_2}$ is allowed to be negative, when $n \in (10, 20)$ and $0 < \rho \leqslant 0.1$, as many as 60% of data sets led to $\widehat{\rho}_{a_1,a_2} < 0$; and we found that negative values of $\widehat{\rho}_{a_1,a_2}$ can lead to large entries in $\widehat{V}^{-1}_{i,a_1,a_2}$ which, in turn, appear to contribute to biased estimates for $\widehat{\Sigma}_{\beta,\eta}$.

### 4.2 Student's t Instead of the Normal Distribution (FSA2)

FSA2 involves using the t distribution (instead of the Normal) with degree-of-freedom equal to $n - p - q$ when constructing confidence intervals ((37)) for $(\beta, \eta)$. Other reference distributions were also considered (e.g., the F distribution with degree-of-freedom equal to $(p+q, n-p-q)$, or using t distribution with Satterthwaite degrees of freedom). In simulations not reported in this paper, however, the t distribution appeared to work best.

### 4.3 Degree-of-freedom Adjustment (FSA3)

FSA3 proposes using $\widehat{\Sigma}^{dof}_{\beta,\eta} = \frac{n}{(n-p-q)}\widehat{\Sigma}_{\beta,\eta}$ in place of $\widehat{\Sigma}_{\beta,\eta}$.

### 4.4 Bias Correction (FSA4)

Let $r_{i,(a_1,a_2)} = \widehat{\epsilon}_{i,(a_1,a_2)}(X_i, Y_i; \hat{\beta}, \hat{\eta})$ denote the estimated residual vector for cluster $i$. Note that the outer product $r_i r_i^T$ is used to compute an estimate of $cov(Y_{i,a_1,a_2})$ within $\widehat{M}$ in $\widehat{\Sigma}_{\beta,\eta}$. In small samples, however, $r_{i,(a_1,a_2)}$ is likely to be closer to zero than the true $\epsilon_{i,(a_1,a_2)}$, often leading to anti-conservative estimates of $\widehat{\Sigma}_{\beta,\eta}$. FSA4 addresses this using an extension of a method first proposed by Mancl & DeRouen ((21)), and subsequently improved on by McCaffrey and Bell ((44; 45)), Tipton, et al ((43)).

The general strategy in Mancl & DeRouen ((21)), as applied to cSMARTs, can be described as having three steps: (i) approximate the bias in $r_{i,(a_1,a_2)}r_{i,(a_1,a_2)}^T$ using a first-order Taylor series expansion around the true residual, i.e. $r_{i,(a_1,a_2)} \approx \epsilon_{i,(a_1,a_2)} + \frac{\partial \epsilon_{i,(a_1,a_2)}}{\partial(\beta,\eta)^T}((\hat{\beta},\hat{\eta}) - (\beta,\eta))$, (ii) approximate $(\hat{\beta},\hat{\eta}) - (\beta,\eta)$ with

$$-\frac{1}{n}\tilde{B}^{-1}(\sum_{i=1}^{n}\sum_{(a_1,a_2)} I_{i,a_1,a_2}(S_i) W_{i,a_1,a_2}(X_i,S_i) D_{a_1,a_2}(X_i)^T V_{i,a_1,a_2}^{-1}(X_i,\alpha) \epsilon_{i,(a_1,a_2)})).$$

using a Taylor series expansion of $\mathbb{P}_n U_{\hat{\beta},\hat{\eta}} = 0$ around $(\beta,\eta)$; and (iii) combine these two approximations to calculate a *bias-corrected* estimate of $\widehat{\Sigma}_{\beta,\eta}$ (see Appendix 2).

However, there is a challenge to directly incorporating this strategy with data from a cSMART: Some of the clusters $i$ have observed data that is consistent with more than one of the embedded cAIs (e.g., responding schools in the ASIC SMART), making $U_{i,\beta,\eta}$ the sum of a function of *multiple residuals*. This property of the estimator—which results from the combination of the sum over the embedded cAIs and the inside indicator $I_{i,a_1,a_2}(S_i)$ (see Display 3)—leads to an abundance of cross-product terms $r_{i,(a_1,a_2)}r_{i,(a_1',a_2')}^T$ in $M = U_{i,\beta,\eta}U_{i,\beta,\eta}^T$ and $\epsilon_{i,(a_1,a_2)}\epsilon_{i,(a_1',a_2')}^T$. (In the case of the prototypical cSMART with 4 embedded cAIs, there are a total of 6 cross-product terms.) This challenge is unique to cSMARTs in that these cross-product terms typically do not arise when analyzing data from standard, clustered, confirmatory randomized trials.

Following the lead of McCaffrey and Bell ((44; 45)), we sought to overcome this challenge by finding a matrix $Q_{i,a_1,a_2}$ which ensures that $E((\sum_{(a_1,a_2)} I_{i,a_1,a_2} W_{i,a_1,a_2}(X_i,S_i) Q_{i,a_1,a_2} r_{i,(a_1,a_2)})^{\otimes 2})$ is, on average, close to $E((\sum_{(a_1,a_2)} I_{i,a_1,a_2} W_{i,a_1,a_2}(X_i,S_i) D_{a_1,a_2}(X_i)^T V_{i,a_1,a_2}^{-1}(X_i) \epsilon_{i,(a_1,a_2)})^{\otimes 2})$. Appendix 2 describes the algebraic calculations used to identify such a matrix $Q_{i,a_1,a_2}$. Additional algebra shows that the proposed FSA amounts to replacing $M$ in Display 5 with $\widetilde{M} = E(\widetilde{U}_{i,\beta,\eta}\widetilde{U}_{i,\beta,\eta}^T)$ where

$$\widetilde{U}_{i,\beta,\eta} = (I_{q+p} - \sum_{(a_1,a_2)} I_{i,a_1,a_2} W_{i,a_1,a_2}(X_i,S_i) D_{a_1,a_2}(X_i)^T V_{i,a_1,a_2}^{-1}(X_i) D_{a_1,a_2}(X_i) (nB)^{-1})^{-1} U_{i,\beta,\eta}.$$

4.5 *R Code*

R code that allows users to select any combination of the FSAs described in Sections 3 and 4 is available at `https://github.com/panwenchu98/clusterSMART`.

## 5. Simulations

This section reports the performance of FSA 1 (minimal adjustment) vs FSA 1+2+4 (complete). As expected, results show that FSA1 led to the least favorable coverage rates, whereas FSA 1+2+4 led to the most favorable coverage rates. All other FSAs, used singly or in combination, performed better than FSA1 but under-performed relative to FSA 1+2+4.

5.1 *Data-generative Models*

Pseudo-algorithm 2 shows the steps used to generate the data. For each cluster $i \in (1, \ldots, n)$ of size $m_i$, this approach generates observations $O_i = (X_i, A_{1i}, R_i, A_{2i}, Y_i)$. Each element in $(X_i, A_{1i}, R_i, A_{2i})$ is a scalar; and $Y_i$ is an $m_i \times 1$-dimensional vector of the research outcomes. Data is generated sequentially, from each of the conditional probability distributions $f(\cdot)$ that factor the joint distribution of $O$ in temporal order, as follows: $f(X) \times f(A_1 \mid X) \times f(R \mid X, A_1) \times f(A_2 \mid X, A_1, R) \times f(Y \mid X, A_1, R, A_2)$.

Simulated data sets are manipulated along five dimensions: (1) *Sample size*. We will vary $n$ and $m_i$ (for each $i$). (2) *Effect size*. Without loss of generality, simulations focus on $\Delta_{\left[\begin{smallmatrix}(1,1)\\(-1,-1)\end{smallmatrix}\right]} = E(\mu(1,1,X) - \mu(-1,-1,X))$ with true value 3.5. To study performance across different magnitudes, we will vary the standardized effect size $\delta = 3.5/\text{sd}(Y)$, where $\text{sd}(Y)$ is the standard deviation of the observed $Y$, marginal over $(X, A_1, R, A_2)$; $\delta$ values of 0.2, 0.5, and 0.8 are considered small, moderate and large, respectively (50). Given their roles in sample size and statistical efficiency considerations (9), simulations also will vary the: (3) *Intracluster Correlation Coefficient* (ICC) for $Y$, marginal over $(X, A_1, R, A_2)$ (note the ICC can range from 0 to 1, but most cluster-randomized trials have an ICC in the range of 0.01 to 0.2 (CITE)); (4) *Response Rates* $\tilde{\kappa}(a_1) = Pr(R(a_1) = 1)$; and the (5) *Correlation between $X$ and $Y$*, $\text{cor}(X, Y)$, marginal over the observed $(A_1, R, A_2)$. For each combination of inputs, 10,000 data sets were generated.

In Step 5, $Y$ is generated using $f(Y \mid X, A_1, R, A_2)$ (which is equivalent to $f(Y(a_1, a_2) \mid X, R(a_1))$ under consistency and sequential ignorability). To do this based on the marginal distribution $f(Y(a_1, a_2) \mid X)$ (which includes the target mean model in (2)), we use:

$$Y_i(a_1, a_2) = \xi(X, a_1, R(a_1), a_2) + e_{(a_1, a_2)}(X, R(a_1))$$

$$= \mu(a_1, a_2, X) + \{\xi_{a_1, a_2}(X, R(a_1)) - \mu(a_1, a_2, X) + e_{(a_1, a_2)}(X, R(a_1))\}$$

$$= \mu(a_1, a_2, X) + \epsilon_{(a_1, a_2)}(X), \quad (8)$$

where $E(\epsilon_{(a_1, a_2)(X)}) = 0)$, $E(e_{(a_1, a_2)(X, R(a_1))}) = 0$, and $E(\xi_{a_1, a_2}(X, R(a_1)) \mid X) = \mu(a_1, a_2, X)$ where the third expectation is over the distribution $f(R(a_1) \mid X)$.

Additional details concerning the data generation process are provided in Appendix X.

---

**Pseudo-algorithm 2** Approach to generating data for the simulation experiments.

---

**Input** $f(X)$, $f(Y(a_1, a_2) \mid X)$, and $f(R(a_1) \mid X)$  where $X$ is standard normal

    Sample size: Number of clusters $n$; and cluster sizes $m_1, \ldots, m_n$
    Mean:  $\mu(a_1, a_2, X) = E(Y(a_1, a_2) \mid X)$
        [ includes causal effects and $\eta / \text{sd}(Y) = \text{cor}(X, Y)$ ]
    Var:  $V(a_1, a_2, X) = \text{var}(Y(a_1, a_2) - \mu(a_1, a_2) \mid X) = \text{var}(\epsilon_{(a_1, a_2)} \mid X)$
        [ includes ICC ]
    Response Rate:  $\kappa(a_1, X) = Pr(R(a_1) = 1 \mid X) \quad \Rightarrow \quad \tilde{\kappa}(a_1) = E(\kappa(a_1, X))$

**Calculate** $f(Y(a_1, a_2) \mid X, R(a_1))$

    Mean: $\xi_{a_1, a_2}(X, R(a_1)) = E(Y(a_1, a_2) \mid X, R(a_1))$
        [ includes stage-specific conditional causal effects and $\text{cor}(R(a_1), Y(a_1, a_2))$ ]
    Var: $\Upsilon_{a_1, a_2}(X, R(a_1)) = \text{var}(Y(a_1, a_2) - \xi_{a_1, a_2}) \mid X, R(a_1)) = \text{var}(e_{(a_1, a_2)} \mid X, R(a_1))$

**Generate**

    Step 1 [$X$]:  Generate $X \sim N_n(0, 1)$
    Step 2 [$A_1 \mid X$]:  Generate $A_1 \sim 2\,\text{Binomial}(n, 1/2) - 1$
    Step 3 [$R \mid X, A_1$]:  Generate $R \sim \text{Binomial}(n, \kappa(A_1, X))$
    Step 4a [$A_2 \mid X, A_1, R = 0$]:  Generate $A_2 \sim 2\,\text{Binomial}(n - \sum_n R, 1/2) - 1$
    Step 4b [$A_2 \mid X, A_1, R = 1$]:  Set $A_2 =$ n/a for responding clusters (missing by design)
    Step 5 [$Y \mid X, A_1, R, A_2$]:  For each cluster $i$ with data $(X_i, A_{1i}, R_i, A_{2i})$ generate the
        observed research outcome $Y_i \sim N_{m_i}(\xi_{A_1, A_2}(X_i, R_i)\mathbf{1}_{m_i}, \Upsilon_{A_1, A_2}(X_i, R_i))$

5.2 *Experiment*

This experiment compares SE estimation and coverage probabilities using an estimator with: known weights, compound symmetric working variance structure, and covariate adjustment.

Additional experiments are described in Appendix 6 and Supplementary Appendix 2, which study the performance of the FSAs using versions of the estimator that are known to have improved asymptotic efficiency (e.g., using estimated weights).

Table 2 and Figure-2 shows results for data sets across a wide range of sample sizes ($m_i = (3, 10)$, $n = (10, 90)$) with: $\delta = 0.5$ (moderate), ICC $= 0.1$ (moderate), $\tilde{\kappa}(a_1) = 50\%$ (moderate) $\forall a_1$, and cluster-level covariate $X$ satisfying $cov(X, Y) = 0.5$.

We can see from the results that: 1) As expected, When sample size goes large, the estimation for marginal model parameters approaches the true value, and all methods are approaching the truth in terms of variance estimation and coverage rate; 2) In terms of coverage rates, minimal adjustment (FSA1) performs poorly when $n \leqslant 30$, having coverage as low as 75%, and complete adjustment was the best performing (albeit slightly anti-conservative at $n = 10$); 3) In terms of variance estimation, minimal adjustment alone led to underestimation of the variance, whereas the complete adjustment led to overestimation; 4) The size of the cluster, $m$, has minimal influence of the performance, and the number of clusters is dominantly influencing the performance.

Results are similar even when we vary different values of $\delta$, ICC, $\kappa(a_1)$, and $cor(X, Y)$ (see Supplementary Appendix).

Appendix 6 shows the rule of thumbs of utilizing different methods that are known to improve asymptotic efficiency on their finite sample performance.

# 6. Illustrative Data Analysis

To illustrate the impact on statistical inference of using the complete set of FSAs 1+2+3+4 (full adjustment) vs using only FSA1 (minimal adjustment), we use data from the ASIC study, a recent clustered SMART described in Section **??** ((11; 40)).

## 6.1 *Focal Estimands*

We focus on two marginal mean comparisons on the outcome $Y_{ij}$, the average number of CBT sessions delivered by SP's within a school in Phase 3 of implementation (from week 33 to week 43); more CBT delivery is considered better. Estimand 1 is $\Delta_{\begin{bmatrix}(1,1)\\(-1,-1)\end{bmatrix}}$, which compares REP+COACH+FAC vs REP Only. Estimand 2 is $\Delta_{\begin{bmatrix}(1,-1)\\(-1,1)\end{bmatrix}}$, which compares REP+COACH vs REP+FAC. These two estimands were selected because they differed significantly in terms of their estimated effect sizes (and inference) as reported in the ASIC's primary manuscript ((11)).

Estimand 1 will illustrate the impact of the FSA's for ASIC's primary aim comparison, for which there was no evidence of a clinically significant effect. Whereas, Estimand 2 will illustrate the impact of the FSA's for the comparison that led to the largest estimated pairwise causal effect. Note the estimated rank-ordering of the cAIs in terms of $\widehat{E}(Y(a_1, a_2))$ was: REP+FAC < REP ≈ REP+COACH+FAC < REP+COACH. Two additional features make these comparisons interesting: First, each of these estimands is a comparison of a cAI versus a non-adaptive clustered intervention. Second, the results were unexpected for both– in the design of the trial, it was hypothesized that there would be a significant difference in the Estimand 1 comparison, and we did not hypothesize a difference in the Estimand 2 comparison.

For completeness, in Supplementary Appendix X we provide results for a variety of other causal estimands.

6.2 *Modeling*

All analyses are based on the marginal structural model in display (2) with the following six baseline covariates in $X$: school size ($\geqslant 500$ or $< 500$ students; `large`), location of school (rural or urban; `rural`), percentage of students on free/reduced lunch program ($\geqslant 50\%$ or $< 50\%$; `pctFR`), and phase 1/pre-randomization CBT delivery (any vs. none; `anyCBT`), school-aggregated SP education (school-level proportion of SPs with graduate degree; `educ`) and job tenure (years; `tenure`). These covariates were pre-specified for use in ASIC's primary aim analyses ((11; 40)). All analyses use a working compound symmetric model with heterogeneous total variance—i.e., using the working model $\sigma^2_{a_1,a_2} CS(\rho)$ for $V_{i,a_1,a_2}(X_i)$.

6.3 *Analytic Plan*

Two sets of illustrative analyses were planned. In the first set, we planned to analyze the original ASIC study data. Given the results of the simulation experiments, we hypothesized that: (i) for both estimands, there would be no or minimum differences in statistical inference between the full vs minimal FSAs for a prototypical cSMART as large as ASIC, i.e., with $N = 94$, $\sum_i^N m_i = 192$; and (ii) these results would be very similar compared to inferences made in ASIC's primary aims manuscript.

In a second set of analyses, we planned to re-analyze the data *had ASIC been designed using progressively smaller sample sizes*. To do this, four counterfactual ASIC studies were "created" by drawing random samples of the 94 schools in ASIC, each of size $N_\kappa = \kappa\, 94$, where $\kappa = 2/3, 1/2, 1/3, 1/4$, respectively. For each $\kappa$, $B = 1000$ random samples were drawn without replacement (enabling us to understand the operating characteristics of the estimators for the possible range of "ASIC studies" of size $N_\kappa$). To prevent counterfactual data sets with empty sequences $S = (A_1, R, A_2)$, each random sample of schools was stratified by the six unique values of $S$.

6.4 *Hypotheses*

In the simulation experiments, we observed that in the absence of FSAs, there is significant underestimation of $\widehat{\text{var}}(\hat{\theta})$ for cSMARTs with progressively smaller sample sizes; and this is more likely to have a negative impact on statistical inference for estimands with smaller effect sizes. Thus, our hypothesis for Estimand 1 ($\Delta_{\left[\begin{smallmatrix}(1,1)\\(-1,-1)\end{smallmatrix}\right]}$), which had an estimated effect size near zero, was that we would begin to see differences in statistical inference resulting from using the full vs minimal FSAs for smaller-sized ASIC studies. Specifically, we hypothesized that for smaller values of $\kappa$, we would observe: (i) greater differences in the proportion of 95% confidence intervals not covering zero, i.e., $\texttt{nozero}_{\min} - \texttt{nozero}_{\text{full}}$, where $\texttt{nozero} = \mathbb{P}_B(0 \notin 95\%\text{CI}(\widehat{\Delta}))$; (here subscript min represents using only the minimal adjustment, and full represents using all adjustments) and (ii) larger values in the ratio of standard error estimation comparing with FSA to without FSA, i.e., $\texttt{bootse}_{\min}/\texttt{bootse}_{\text{full}} < 1$, where $\texttt{bootse} = \sqrt{\mathbb{P}_B(\widehat{\text{var}}(\widehat{\Delta}))}$ is the square root of the bootstrapped mean of variance estimation.

Since Estimand 2 ($\Delta_{\left[\begin{smallmatrix}(1,-1)\\(-1,1)\end{smallmatrix}\right]}$) had a larger estimated effect size and would be considered significant by using full data analysis, we hypothesized that, for smaller values of $\kappa$, (i) $\texttt{nozero}_{\min} - \texttt{nozero}_{\text{full}}$ would change less noticeably, if at all; yet (ii) we expect $\texttt{bootse}_{\min}/\texttt{bootse}_{\text{full}}$ to increase as it did with Estimand 1.

6.5 *Results*

Table 3 illustrates the analysis result of the primary aim analysis of full ASIC data. From the result we can see: (i) there is less than 10% difference in the standard error estimation comparing using minimal FSA with proposed FSA, thus all but one coefficient keep their (in)significance; (ii) by comparing with the primary aim result paper ((40)), the outcomes are similar.

Table 4 illustrates the analysis result comparing the metrics of the two interested estimands. From the table we can see: For Estimand 1 which is small in effect size and not significant, (i) inference with $\kappa = 2/3$ has enough sample size to guarantee non-significance, while smaller $\kappa$ will lead to false significance, and the difference in false significant rate comparing proposed FSA with minimal FSA increases with smaller $\kappa$; (ii) the ratio of standard error estimation increases with smaller $\kappa$. For Estimand 2 which has the largest effect size and is significant in inference with full data, (i) decrease in sample size leads to less certainty of claiming significance, and the differnce in significant rate comparing proposed FSA with minimal FSA keeps stable; (ii) the ratio of standard error estimation increases with smaller $\kappa$.

Summarizing the results of two analysis, we can conclude that our proposed FSAs generate reliable results consistent with previous methods, do not introduce bias in large sample size, and have better performance and stability when analyzing data with smaller sample size.

## 7. Discussions and future work

The marginal mean comparison between two (or more) adaptive interventions often represents the target causal estimand(s) in a SMART. For example, a common primary aim comparison in a SMART is to test the null hypothesis that the mean difference between two adaptive interventions is zero ((41; 42)) for a pre-specified primary research outcome/endpoint. The most common estimator for making such comparisons is an inverse-probability-of-intervention weighted least squares regression approach, with weights that are a function of the known randomization probabilities ((17; 1)).

In a clustered SMART (cSMART; (9; 31; 32; 33)), the sequence of randomizations are at the level of intact groups of units (e.g., schools are (re)randomized), yet the primary outcome is at the level of a nested group of units (e.g., school professionals within each school). As with standard multi-arm clustered randomized trials, standard methods for statistical inference work well for a large number of clusters, but fail for cSMARTs with small sample sizes.

This manuscript presents, evaluates and illustrates four finite sample adjustment (FSA) methods recommended for use when comparing the embedded adaptive interventions in a clustered, sequential, multiple assignment randomized trials (cSMART). To our knowledge, this is the first biostatistical manuscript to propose any FSA for these comparisons using data from a SMART.

Based on the results of an extensive set of simulation experiments, conducted across a wide variety of data generative models and sample sizes, the combination of all four FSA's performs well in terms of coverage rate of confidence interval and estimation of variance. Using the four FSA's always outperforms not using any FSA. And the use of all four FSA's provides adequate coverage rate and precision of variance estimation for clustered SMARTs as small as N=10 clusters regardless of the effect sizes of the interested marginal comparison (effect size = 0.2, 0.5, 0.8) or the cluster-level covariates (effect size= 0, 0.2, 0.5, 0.8 (when main effect size=0.5) or cluster size (vary from 1-3 to 10-30).

Admittedly, there are many other techniques and arguments on the modification of the original robust sandwich estimator to improve its finite sample performance, based on different assumptions. Wei Pan proposed to use pooled covariance estimation based on all subjects((23)). Lipschitz proposed an one-step Jackknife estimator asymptotically equivalent to the original sandwich estimator but performs better on finite sample setting((24)). Morel proposed an estimator to account for the possible over-dispersion((22)). And Westgate took into account the change in variance due to estimating the correlation structure((25)). These methods all generate promising results on improving the performance of variance estimator in clustered GEE, but some of their assumptions are violated in cSMART, so we leave the adaptation of these estimators for future work.

The general direction for possible future work is to include more information or use more flexible modeling in order to improve efficiency. The first possible direction is utilizing more information in estimating the inverse probability weights. Currently we use only empirical weights, but the underlying truth maybe the randomization of decision is subject to other factors like professional recommendations and patient preference, so using a parametric model with baseline covariates can help to gain efficiency. Another possible direction is utilizing longitudinal data (e.g. in ASIC dataset there are longitudinal evaluations for each school professional) and introducing multi-level structure in the model. Another direction is extension of such method to SMART studies with more stages of randomization or more levels of nested clustering, to meet up with the evolving needs in the field. Another possible direction is to utilize more flexible choice of modeling, like structural nested mean model and other semiparametric methods, in order to gain efficiency.

## Acknowledgements

The authors are grateful for the work of Shawna Smith, Elizabeth Koschmann and others on the ASIC Study Team, in conducting the trial and collected the data. The authors are also grateful for the fellow colleagues at d3c center, including Gabriel Durham, Mason Ferlic, Timothy Lycurgus for their help and insight during the completion of the paper. The work is sponsored by grant P50DA054039, R01DA047279 and R01DA039901.

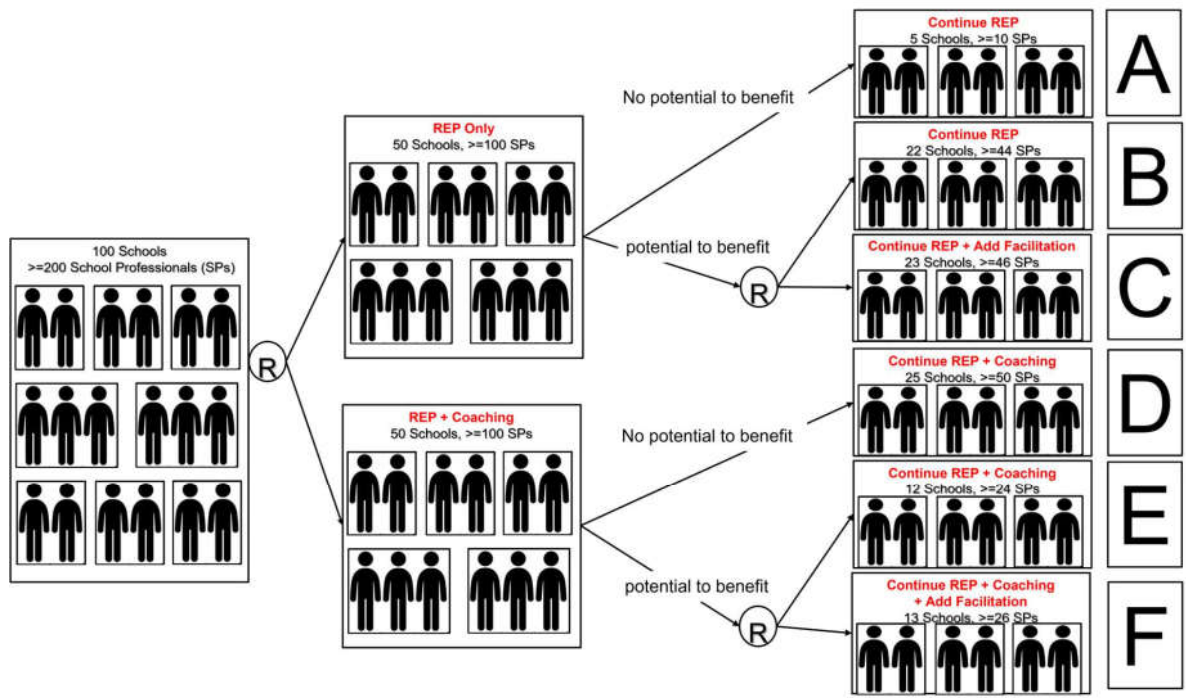

**Figure 1**. Schematic of ASIC SMART trial

**REP**: Replicating Effective Programs; low-level implementation strategy that provides manualization of intervention (e.g., CBT), didactic training, & technical assistance.

**Coaching**: In-person coaching during CBT groups at the school for a minimum 12 weeks.

**Facilication**: Phone calls with an expert in CBT & strategic thinking for a minimum 12 weeks.

**Response status** at the end of the first-stage is being estimated to have no potential to benefit from further facilitation. The potential of benefit is based on a combination of two measures:

(i) >= 1 participating School Professionals delivering < 3 cognitive behavioral therapy (CBT) components to < 10 students, or

(ii) school professionals (SPs) reporting, on average, > 2 barriers to CBT uptake

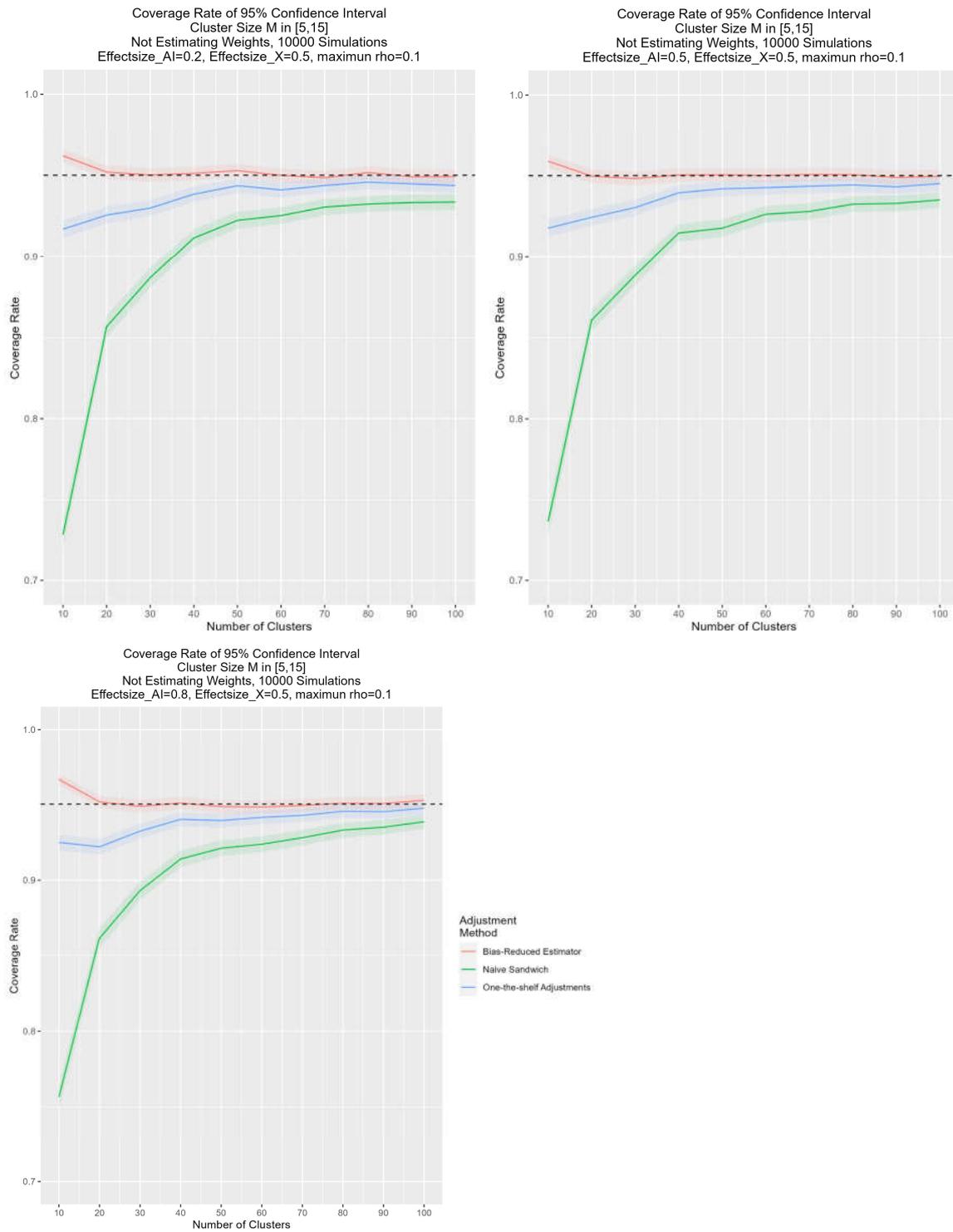

**Figure 2.** Coverage Rate of confidence interval generated by variance estimator using: minimal FSA, on-the-shelf FSA, and proposed FSA

Table 1: The Four interventions embedded in the ASIC study.

| AI($a_1, a_2$) | First-Stage Intervention | School Response Status | Second-Stage Intervention | Cells in Figure |
|---|---|---|---|---|
| AI(1,1) | REP Only ($A_1$=1) | No Potential (R=0) | Continue REP | A |
| | | Have Potential (R=1) | Continue REP ($A_2$=1) | B |
| AI(1,-1) | REP Only ($A_1$=1) | No Potential (R=0) | Continue REP | A |
| | | Have Potential (R=1) | Continue REP + Add Facilitation ($A_2$=-1) | C |
| AI(-1,1) | REP + Coaching ($A_1$=-1) | No Potential (R=0) | Continue REP + Coaching | D |
| | | Have Potential (R=1) | Continue REP + Coaching ($A_2$=1) | E |
| AI(-1,-1) | REP + Coaching ($A_1$=-1) | No Potential (R=0) | Continue REP + Coaching | D |
| | | Have Potential (R=1) | Continue REP +Coaching + Add Facilitation ($A_2$=-1) | F |

Table 2: Simulation to illustrate finite-sample inference challenges ($\delta = 0.5$, ICC= 0.2).

| Sample Size[a] | | | Causal Effect $\Delta$ Estimation[b] | | | | Standard Error Estimation[c] | | 95% Confidence Interval Coverage Rate[d] | |
|---|---|---|---|---|---|---|---|---|---|---|
| $m$ | $n$ | $N$ | avg($\widehat{\Delta}$) | Bias | sd($\widehat{\Delta}$) | RMSE | Unadjusted | Finite Sample Adjusted | Unadjusted | Finite Sample Adjusted |
| 5 | 10 | 50 | 3.317 | -0.183 | 4.848 | 4.851 | 3.382 | 6.849 | 0.740 | 0.964 |
|  | 20 | 100 | 3.399 | -0.101 | 3.307 | 3.309 | 2.818 | 3.885 | 0.859 | 0.952 |
|  | 30 | 150 | 3.403 | -0.097 | 2.654 | 2.656 | 2.386 | 2.879 | 0.894 | 0.950 |
|  | 50 | 250 | 3.474 | -0.026 | 2.007 | 2.007 | 1.895 | 2.100 | 0.921 | 0.950 |
|  | 70 | 350 | 3.460 | -0.040 | 1.661 | 1.662 | 1.607 | 1.723 | 0.931 | 0.950 |
|  | 90 | 450 | 3.435 | -0.065 | 1.481 | 1.483 | 1.421 | 1.499 | 0.935 | 0.950 |
| 10 | 10 | 100 | 3.265 | -0.235 | 4.712 | 4.717 | 3.283 | 6.677 | 0.742 | 0.964 |
|  | 20 | 200 | 3.343 | -0.157 | 3.253 | 3.257 | 2.754 | 3.803 | 0.857 | 0.948 |
|  | 30 | 300 | 3.462 | -0.038 | 2.566 | 2.567 | 2.352 | 2.841 | 0.902 | 0.955 |
|  | 50 | 500 | 3.458 | -0.042 | 1.963 | 1.963 | 1.860 | 12.061 | 0.921 | 0.952 |
|  | 70 | 700 | 3.510 | 0.010 | 1.654 | 1.654 | 1.583 | 1.698 | 0.927 | 0.948 |
|  | 90 | 900 | 3.495 | -0.005 | 1.448 | 1.448 | 1.402 | 1.478 | 0.931 | 0.948 |

[a] $m = m_i$ for all clusters $i = 1, \ldots, n$; and the total sample size is $N = \sum_i^n m_i = n \times m$.
[b] $\Delta_{true} = E(Y(1,1) - Y(-1,-1)) = 3.5$ with a moderate, standardized effect size of $\delta = 1/2$.
  avg($\widehat{\Delta}$) is the average value of $\widehat{\Delta}$ across the 10,000 simulated data sets.
  Bias is defined as avg($\widehat{\Delta}$) $- \Delta_{true}$.
  sd($\widehat{\Delta}$) is the standard deviation of $\widehat{\Delta}$ across the 10,000 simulated data sets; i.e., an estimate of $\widehat{\Delta}$'s true standard error.
  The root mean squared error (RMSE) is defined as $(\text{avg}(\widehat{\Delta} - \Delta_{true})^2)^{1/2}$.
[c] Unadjusted SE's are calculated using the formulae in Displays (4) and (5) in Section 3.
  Finite Sample Adjusted SE's are calculated using FSA 1, 3, 4 described in Section 4.
[d] Coverage Rate is calculated as the percentage of the 95% confidence intervals (out of 10,000) that include the $\Delta_{true}$.
  Unadjusted CIs uses the unadjusted SEs described in Section 3 and confidence levels based on the Normal distribution.
  Finite Sample Adjusted CIs are calculated using all four of the FSAs (1-4) described in Section 4.

Table 3: Coverage rate of 95% confidence interval under proposed finite-sample adjustments

| n | $\delta = 0.2$ | | | $\delta = 0.5$ | | | $\delta = 0.8$ | | |
|---|---|---|---|---|---|---|---|---|---|
|  | minimal FSA | on-the-shelf FSA | proposed FSA | minimal FSA | on-the-shelf FSA | proposed FSA | minimal FSA | on-the-shelf FSA | proposed FSA |
| 10 | 0.728 | 0.917 | 0.964 | 0.736 | 0.918 | 0.963 | 0.757 | 0.925 | 0.969 |
| 20 | 0.857 | 0.926 | 0.953 | 0.861 | 0.924 | 0.951 | 0.862 | 0.922 | 0.953 |
| 30 | 0.887 | 0.930 | 0.950 | 0.889 | 0.930 | 0.948 | 0.893 | 0.932 | 0.950 |
| 50 | 0.922 | 0.944 | 0.953 | 0.918 | 0.942 | 0.950 | 0.921 | 0.940 | 0.949 |
| 70 | 0.931 | 0.944 | 0.949 | 0.928 | 0.944 | 0.951 | 0.928 | 0.943 | 0.949 |
| 90 | 0.933 | 0.945 | 0.947 | 0.933 | 0.943 | 0.949 | 0.925 | 0.945 | 0.951 |

[a] Adjustment Method: 1 for forcing non-negative ICC, 2 for using t distribution, 3 for degree-of-freedom adjustment, 4 for using bias-reduction.
[b] minimal FSA indicates using FSA 1 only, On-the-shelf FSA indicates using FSA 1+2+3, proposed FSA indicates using FSA 1+2+4

Table 4: Analysis of the complete ASIC study data.

| | | | With FSA 1 only, 95% Confidence | | | | With FSAs 1-4, 95% Confidence | | | |
|---|---|---|---|---|---|---|---|---|---|---|
| | | EST | SE | CL-low | CL-upp | p-value | SE | CL-low | CL-upp | p-value |
| Causal Terms | intercept | 31.492 | 3.036 | 25.542 | 37.442 | **<1e-20** | 3.267 | 24.994 | 37.989 | **3.1e-15** |
| | $a_1$ | -4.818 | 3.046 | -10.788 | 1.153 | 0.114 | 3.284 | -11.348 | 1.712 | 0.146 |
| | $a_2$ | 5.690 | 2.791 | 0.220 | 11.160 | **0.041** | 3.019 | -0.314 | 11.694 | 0.063 |
| | $a_1 : a_2$ | -2.104 | 2.764 | -7.520 | 3.313 | 0.447 | 2.994 | -8.057 | 3.850 | 0.484 |
| Covariate Terms | large | 4.931 | 7.048 | -8.883 | 18.745 | 0.484 | 7.674 | -10.329 | 20.191 | 0.522 |
| | pctFR | -5.205 | 5.904 | -16.776 | 6.365 | 0.378 | 6.424 | -17.979 | 7.569 | 0.420 |
| | rural | 2.888 | 6.808 | -10.456 | 16.233 | 0.671 | 7.434 | -11.895 | 17.672 | 0.699 |
| | anyCBT | 4.603 | 5.970 | -7.099 | 16.305 | 0.441 | 6.507 | -8.337 | 17.543 | 0.481 |
| | educ | 8.718 | 10.050 | -10.980 | 28.417 | 0.386 | 11.060 | -13.275 | 30.711 | 0.433 |
| | tenure | 0.152 | 0.445 | -0.721 | 1.025 | 0.733 | 0.484 | -0.811 | 1.115 | 0.754 |
| Effects | $\Delta_{\left[\begin{smallmatrix}(1,1)\\(-1,-1)\end{smallmatrix}\right]}$ | 1.744 | 7.724 | -13.394 | 16.883 | 0.821 | 8.427 | -15.014 | 18.503 | 0.837 |
| | $\Delta_{\left[\begin{smallmatrix}(1,-1)\\(-1,1)\end{smallmatrix}\right]}$ | -21.016 | 8.769 | -38.202 | -3.830 | **0.002** | 9.389 | -39.688 | -2.345 | **0.003** |
| | $\Delta_{\left[\begin{smallmatrix}(1,1)\\(1,-1)\end{smallmatrix}\right]}$ | 7.173 | 6.942 | -6.433 | 20.779 | 0.301 | 7.588 | -7.917 | 22.263 | 0.347 |
| | $\Delta_{\left[\begin{smallmatrix}(1,1)\\(-1,1)\end{smallmatrix}\right]}$ | -13.843 | 9.388 | -32.244 | 4.557 | 0.140 | 10.098 | -33.924 | 6.237 | 0.174 |
| | $\Delta_{\left[\begin{smallmatrix}(1,-1)\\(-1,-1)\end{smallmatrix}\right]}$ | -5.429 | 6.870 | -18.894 | 8.037 | 0.429 | 7.482 | -20.308 | 9.451 | 0.470 |
| | $\Delta_{\left[\begin{smallmatrix}(-1,1)\\(-1,-1)\end{smallmatrix}\right]}$ | 15.588 | 8.673 | -1.410 | 32.586 | 0.072 | 9.330 | -2.965 | 34.141 | 0.98 |

Table 5: Analysis of progressively smaller-sized ASIC study datasets to illustrate

| | Focal Estimand 1 $\Delta_{\left[\begin{smallmatrix}(1,1)\\(-1,-1)\end{smallmatrix}\right]}$ | | | | Focal Estimand 2 $\Delta_{\left[\begin{smallmatrix}(1,-1)\\(-1,1)\end{smallmatrix}\right]}$ | | | |
|---|---|---|---|---|---|---|---|---|
| $\kappa$ | nozero$_{\text{min}}$ | nozero$_{\text{full}}$ | nozero$_{\text{min}}$ − nozero$_{\text{full}}$ | $\frac{\text{bootse}_{\text{min}}}{\text{bootse}_{\text{full}}}$ | nozero$_{\text{min}}$ | nozero$_{\text{full}}$ | nozero$_{\text{min}}$ − nozero$_{\text{full}}$ | $\frac{\text{bootse}_{\text{min}}}{\text{bootse}_{\text{full}}}$ |
| 2/3 | 0.0 | 0.0 | 0.0 | 1.129 | 52.5 | 32.0 | 20.5 | 1.108 |
| 1/2 | 1.6 | 0.4 | 1.2 | 1.177 | 37.5 | 18.2 | 19.3 | 1.157 |
| 1/3 | 3.8 | 0.6 | 3.2 | 1.264 | 24.9 | 9.2 | 15.7 | 1.245 |
| 1/4 | 8.0 | 1.6 | 6.4 | 1.404 | 21.7 | 5.4 | 16.3 | 1.386 |

[a] nozero $= \mathbb{P}_B(0 \notin 95\%\text{CI}(\widehat{\Delta}))$ is the proportion of 95% confidence interval not covering zero.
[b] subscript min means using minimal FSA (FSA 1 only) and subscript full means using all FSA (FSA 1-4).
[c] bootse $= \sqrt{\mathbb{P}_B(\widehat{\text{var}}(\widehat{\Delta}))}$ is the square root of the bootstrapped mean of variance estimation.

# A

## A.1 Appendix 1. Justification of Constrained Maximum Likelihood Estimator

In this part, we are going to prove that the solution from the GEE equation and our proposed method generates a constrained maximum likelihood estimator for $\beta_i$, $\eta$, $\sigma^2_{(a_1,a_2)}$ and $\rho_{(a_1,a_2)}$ if marginal exchangeable assumption holds.

We start from the most trivial setting, where only 1 Dynamic Treatment Regime is applied to all clusters, all clusters are of the same size m, there is only one-dimensional cluster-level covairate, and marginal exchangeable assumption holds.

That means $\mu_i = \mu$, $\sigma_i^2 = \sigma^2$ and $\rho_i = \rho$,

$Y_i = (Y_{i1}, \ldots, Y_{im}|^T = \mu \cdot \mathbf{1}_m + \eta X_i \cdot \mathbf{1}_m + \epsilon_i$, where $\epsilon_i = (\epsilon_{i1}, \ldots, \epsilon_{im})^T \sim N_m(0_m, \Sigma_i = \sigma^2 CS_m(\rho))$

Under this setting, the GEE equation becomes

$$0 = \sum_{i=1}^{n} U_i(X_i, Y_i; \beta, \eta| \triangleq \sum_{i=1}^{n} D(X_i|^T V_i^{-1}(Y_i - \mu(X_i; \beta, \eta||$$

We notice that, if we use a different setting of parameters to replace $\beta$, say $\widetilde{\beta}$. As long as the new parameters is generated by a non-singular liner transformation of $\beta$, say that $\widetilde{\beta} = A * \beta$, where A is an invertible matrix, then the GEE function becomes

$$0 = \sum_{i=1}^{n} \widetilde{D}(X_i|^T V_i^{-1}(Y_i - \widetilde{\mu}(X_i; \beta, \eta||$$

Because this is just a matter of re-parameterization, $\widetilde{\mu}(X_i; \beta, \eta| = \mu(X_i; \beta, \eta|$. Also, $D(X_i| = \widetilde{D}(X_i| * A$, so the GEE equation becomes $0 = A^{-1} \sum_{i=1}^{n} D(X_i|^T V_i^{-1}(Y_i - \mu(X_i; \beta, \eta||$, which is equivalent to the GEE equation of $\beta$.

So, for simplicity of calculation, we will use $\mu$ instead of $\beta$ in the proof.

Notice that $det(CS_m(\rho)) = (1-\rho)^{m-1} \cdot (1 + (m-1)\rho)$

and $CS_m(\rho)^{-1} = \frac{1+(m-2)\rho}{(1-\rho)(1+(m-1)\rho)} CS_m(-\frac{\rho}{1+(m-2)\rho}) = \frac{1}{1-\rho} I_m - \frac{\rho}{(1-\rho)(1+(m-1)\rho)} J_m$, where $J_m$ is a m*m matrix with 1 in every entry.

Also, now
$$D(X_i|^T = \begin{pmatrix} 1 & 1 & \cdots & 1 \\ X_i & X_i & \cdots & X_i \end{pmatrix}_{m*2}$$

So the GEE equation becomes

$$0 = \sum_{i=1}^n D(X_i|^T V_i^{-1}(Y_i - \mu(X_i; \beta, \eta\|) =$$

$$\sum_{i=1}^n \begin{pmatrix} 1 & 1 & \cdots & 1 \\ X_i & X_i & \cdots & X_i \end{pmatrix} \frac{1}{\sigma^2} CS_m(\rho)^{-1} (Y_i - \mu \cdot \mathbf{1}_m - \eta X_i \cdot \mathbf{1}_m)$$

$$\Leftrightarrow \sum_{i=1}^n \begin{pmatrix} 1 \\ X_i \end{pmatrix} \mathbf{1}_m^T \left(I_m - \frac{\rho}{1+(m-1)\rho} J_m\right) (Y_i - \mu \cdot \mathbf{1}_m - \eta X_i \cdot \mathbf{1}_m) = 0$$

$$\Leftrightarrow \sum_{i=1}^n \begin{pmatrix} 1 \\ X_i \end{pmatrix} \mathbf{1}_m^T (Y_i - \mu \cdot \mathbf{1}_m - \eta X_i \cdot \mathbf{1}_m) = 0$$

$$\Leftrightarrow \sum_{i=1}^n \begin{pmatrix} \sum_{j=1}^m Y_{ij} - m\mu - m\eta X_i \\ X_i \sum_{j=1}^m Y_{ij} - mX_i\mu - mX_i^2\eta \end{pmatrix} = 0$$

These two equations have the unique solution $(\mu, \eta)$

Now we calculate the likelihood function of this model.

$$P(Y_i; \mu, \eta, \sigma^2, \rho) = \frac{1}{(2\pi)^{\frac{m}{2}}|\Sigma|^{\frac{1}{2}}} exp\left\{(Y_i - \mu\mathbf{1}_m - \eta X_i\mathbf{1}_m)^T \Sigma^{-1}(Y_i - \mu\mathbf{1}_m - \eta X_i\mathbf{1}_m)\right\}$$

So $P(Y_1, \ldots, Y_n; \mu, \eta, \sigma^2, \rho) = (2\pi)^{-\frac{mn}{2}} \sigma^{-mn} (1-\rho)^{-\frac{n(m-1)}{2}} (1+(m-1)\rho)^{-\frac{n}{2}} *$

$$exp\left\{-\frac{1}{2\sigma^2}\sum_{i=1}^n (Y_i - \mu\mathbf{1}_m - \eta X_i\mathbf{1}_m)^T CS_m(\rho)^{-1}(Y_i - \mu\mathbf{1}_m - \eta X_i\mathbf{1}_m)\right\}$$

Log-likelihood $L(Y_1, \ldots, Y_n; \mu, \eta, \sigma^2, \rho) \asymp -\frac{mn}{2}log(\sigma^2) - \frac{n(m-1)}{2}log(1-\rho) - \frac{n}{2}log(1+(m-1)\rho)$

$-\frac{1}{2\sigma^2}\sum_{i=1}^n (Y_i - \mu\mathbf{1}_m - \eta X_i\mathbf{1}_m)^T CS_m(\rho)^{-1}(Y_i - \mu\mathbf{1}_m - \eta X_i\mathbf{1}_m)$

$\frac{\partial L}{\partial \mu} = 0 \Leftrightarrow \frac{\partial \sum_{i=1}^n (Y_i - \mu\mathbf{1}_m - \eta X_i\mathbf{1}_m)^T CS_m(\rho)^{-1}(Y_i - \mu\mathbf{1}_m - \eta X_i\mathbf{1}_m)}{\partial \mu} = 0$

$$\Leftrightarrow \frac{\partial \sum_{i=1}^{n}(-\mu \mathbf{1}_m^T CS_m(\rho)^{-1}(Y_i-\eta X_i \mathbf{1}_m)-(Y_i-\eta X_i \mathbf{1}_m)^T CS_m(\rho)^{-1}\mu \mathbf{1}_m + \mu^2 \mathbf{1}_m^T CS_m(\rho)^{-1}\mathbf{1}_m)}{\partial \mu} = 0$$

$$\Leftrightarrow \sum_{i=1}^{n}(-2\mathbf{1}_m^T CS_m(-\tfrac{\rho}{1+(m-2)\rho})(Y_i - \eta X_i \mathbf{1}_m) + 2\mu \mathbf{1}_m^T CS_m(-\tfrac{\rho}{1+(m-2)\rho})\mathbf{1}_m) = 0$$

$$\Leftrightarrow -(\sum_{i=1}^{n}\sum_{j=1}^{m}(Y_{ij}-\eta X_i)) \cdot 2 \cdot \tfrac{1-\rho}{1+(m-2)\rho} + nm \cdot 2\mu \cdot \tfrac{1-\rho}{1+(m-2)\rho} = 0$$

$$\Leftrightarrow \mu = \frac{\sum_{i=1}^{n}\sum_{j=1}^{m}(Y_{ij}-\eta X_i)}{nm} \Leftrightarrow nm\cdot\mu + m(\sum_{i=1}^{n}X_i)\cdot\eta - \sum_{i=1}^{n}\sum_{j=1}^{m}Y_{ij} = 0$$

Similarly, $\frac{\partial L}{\partial \eta} = 0 \Leftrightarrow \eta = \frac{\sum_{i=1}^{n}(X_i \sum_{j=1}^{m}(Y_{ij}-\mu))}{nm}$

$$\Leftrightarrow m(\sum_{i=1}^{n}X_i)\cdot\mu + m(\sum_{i=1}^{n}X_i^2)\cdot\eta - \sum_{i=1}^{n}\sum_{j=1}^{m}Y_{ij} = 0$$

These two equations are just the same as what we previously derived from the GEE equation. Therefore, the solution of GEE equation is the maximum likelihood estimator of $(\mu, \eta)$

Next, we will show that our proposed formula of $(\sigma^2, \rho)$ is the constrained maximum likelihood estimator.

We define $\Sigma_\mu \triangleq \frac{1}{n}\sum_{i=1}^{n}(Y_i - \mu\mathbf{1}_m - \eta X_i\mathbf{1}_m)(Y_i - \mu\mathbf{1}_m - \eta X_i\mathbf{1}_m)^T$ is the sample covariance matrix of residuals.

$$\frac{\partial L}{\partial \sigma^2} = 0 \Leftrightarrow -\frac{mn}{2\sigma^2} + \frac{1}{2(\sigma^2)^2}\sum_{i=1}^{n}(Y_i-\mu\mathbf{1}_m-\eta X_i\mathbf{1}_m)^T CS_m(\rho)^{-1}(Y_i-\mu\mathbf{1}_m-\eta X_i\mathbf{1}_m) = 0$$

$$\Leftrightarrow \sigma^2 = \frac{1}{mn}\sum_{i=1}^{n}(Y_i-\mu\mathbf{1}_m-\eta X_i\mathbf{1}_m)^T CS_m(\rho)^{-1}(Y_i-\mu\mathbf{1}_m-\eta X_i\mathbf{1}_m)$$

$$= \frac{1}{mn}trace(\sum_{i=1}^{n}(Y_i-\mu\mathbf{1}_m-\eta X_i\mathbf{1}_m)^T CS_m(\rho)^{-1}(Y_i-\mu\mathbf{1}_m-\eta X_i\mathbf{1}_m)) = \frac{1}{m}trace(CS_m(\rho)^{-1}\Sigma_\mu)$$

$$= \frac{1}{m}(\frac{1}{1-\rho}trace(\Sigma_\mu) - \frac{\rho}{(1-\rho)(1+(m-1)\rho)}trace(J_m\Sigma_\mu))$$

$$\frac{\partial L}{\partial \rho} = 0 \Leftrightarrow \frac{n(m-1)}{2}\frac{1}{1-\rho} - \frac{n}{2}\frac{m-1}{1+(m-1)\rho} - \frac{n}{2\sigma^2}\frac{\partial trace(CS_m(\rho)^{-1}\Sigma_\mu)}{\partial \rho} = 0$$

$$\Leftrightarrow \frac{nm(m-1)}{2}\frac{\rho}{(1-\rho)(1+(m-1)\rho)} - \frac{n}{2\sigma^2}\frac{\partial \frac{1}{1-\rho}}{\partial \rho}trace(\Sigma_\mu) + \frac{n}{2\sigma^2}\frac{\partial \frac{\rho}{(1-\rho)(1+(m-1)\rho)}}{\partial \rho}trace(J_m\Sigma_\mu) = 0$$

$\Leftrightarrow \sigma^2 m(m-1)\rho(1-\rho)(1+(m-1)\rho) - (1+(m-1)\rho)^2 trace(\Sigma_\mu) + (1+(m-1)\rho^2)trace(J_m\Sigma_\mu) = 0$

Plug in the above equation for $\sigma^2$, we get

$-(1+(m-1)\rho)trace(\Sigma_\mu) + trace(J_m\Sigma_\mu) = 0 \Leftrightarrow \rho = \frac{1}{m-1}(\frac{trace(J_m\Sigma_\mu)}{trace(\Sigma_\mu)} - 1)$

Plugging back, we get $\sigma^2 = \frac{1}{m}trace(\Sigma_\mu)$

We define $\widehat{\epsilon}_{ij} \triangleq Y_{ij} - \mu - \eta X_i$, then

$trace(\Sigma_\mu) = \frac{1}{n}\sum_{i=1}^{n}\sum_{j=1}^{m}\widehat{\epsilon}_{ij}^2$, and $trace(J_m\Sigma_\mu) = \frac{1}{n}\sum_{i=1}^{n}(\sum_{j=1}^{m}\sum_{k\neq j}^{m}\widehat{\epsilon}_{ij}\widehat{\epsilon}_{ik} + \sum_{j=1}^{m}\widehat{\epsilon}_{ij}^2)$

Thus $\widehat{\sigma}^2 = \frac{\sum_{i=1}^{n}\sum_{j=1}^{m}\widehat{\epsilon}_{ij}^2}{mn}$, and $\widehat{\rho} = \frac{\sum_{i=1}^{n}\sum_{j=1}^{m}\sum_{k\neq j}^{m}\widehat{\epsilon}_{ij}\widehat{\epsilon}_{ik}}{(m-1)\widehat{\sigma}^2}$

This shows our formula calculates the maximum likelihood estimator for $(\sigma^2, \rho)$.

Furthermore, if $\widehat{\rho} < 0$, this is against our constraint of $\rho \in [0,1]$, so the constraint MLE is $\widehat{\rho} = max(\frac{\sum_{i=1}^{n}\sum_{j=1}^{m}\sum_{k\neq j}^{m}\widehat{\epsilon}_{ij}\widehat{\epsilon}_{ik}}{(m-1)\widehat{\sigma}^2}, 0)$

Now we extend the above proof to the SMART setting, where there are 4 dynamic treatment regimens and 6 treatment pathways, and some pathways are shared by two AIs.

For the clusters in shared treatment pathways, we no longer use $P(Y_i; \mu, \eta, \sigma^2, \rho)$. Rather, we split it into $P(Y_i; \mu, \eta, \sigma^2, \rho, a_1, a_2)^\lambda P(Y_i; \mu, \eta, \sigma^2, \rho, a_1, \bar{a}_2)^{1-\lambda}$. By taking logarithm, this becomes the inverse probability weighting, and

$$L(Y_1, \ldots, Y_n) = \sum_{i=1}^{n} \sum_{(a_1,a_2)} I_{i,a_1,a_2} W_i L(Y_i; \mu_{a_1,a_2}, \sigma^2_{a_1,a_2}, \rho_{a_1,a_2})$$
$$= \sum_{(a_1,a_2)} (\sum_{i=1}^{n} I_{i,a_1,a_2} W_i L(Y_i; \mu_{a_1,a_2}, \sigma^2_{a_1,a_2}, \rho_{a_1,a_2}))$$

The parameters of each AI are separated, so our previous proof holds for every AI, thus for the whole algorithm.

## A.2 Appendix 2. Methodology derivation of sandwich variance estimators

For Liang & Zeger estimator,

We denote $\theta = (\beta, \eta)$ as all the parameters in the marginal mean model, then our GEE equation is

$$0 = \sum_{i=1}^{n} U_{i,\widehat{\beta},\widehat{\eta}} \triangleq \sum_{i=1}^{n} \sum_{(a_1,a_2)} I_{i,a_1,a_2}(S_i) W_{i,a_1,a_2}(X_i, S_i) D_{a_1,a_2}(X_i)^T V^{-1}_{i,a_1,a_2}(X_i, \widehat{\alpha})(Y_i - \mu(a_1, a_2, X_i; \widehat{\theta}))$$
$$= \sum_{i=1}^{n} \sum_{(a_1,a_2)} I_{i,a_1,a_2}(S_i) W_{i,a_1,a_2}(X_i, S_i) D_{a_1,a_2}(X_i)^T V^{-1}_{i,a_1,a_2}(X_i, \alpha)(Y_i - \mu(a_1, a_2, X_i; \theta)) -$$
$$(\sum_{i=1}^{n} \sum_{(a_1,a_2)} I_{i,a_1,a_2}(S_i) W_{i,a_1,a_2}(X_i, S_i) D_{a_1,a_2}(X_i)^T V^{-1}_{i,a_1,a_2}(X_i, \widetilde{\alpha}) D_{a_1,a_2}(X_i))(\widehat{\theta} - \theta) + o(1)$$

$$\Leftrightarrow \sqrt{n}(\widehat{\theta}-\theta) = (\sum_{i=1}^{n} \sum_{(a_1,a_2)} I_{i,a_1,a_2}(S_i)W_{i,a_1,a_2}(X_i,S_i)D_{a_1,a_2}(X_i)^T V_{i,a_1,a_2}^{-1}(X_i,\widetilde{\alpha})D_{a_1,a_2}(X_i))^{-1} \cdot$$

$$(\sqrt{n}\sum_{i=1}^{n} \sum_{(a_1,a_2)} I_{i,a_1,a_2}(S_i)W_{i,a_1,a_2}(X_i,S_i)D_{a_1,a_2}(X_i)^T V_{i,a_1,a_2}^{-1}(X_i,\alpha)(Y_i-\mu(a_1,a_2,X_i;\theta)))+$$

$o(1)$

Here $\widetilde{\alpha}$ is somewhere between $\widehat{\alpha}$ and $\alpha$, and by consistency of $\widehat{\theta}$, we have $\widetilde{\theta} \to \theta$ as $n \to \infty$, therefore by calculation of $\widehat{\alpha}$, we have $\widetilde{\alpha} \to \alpha$.

By central limit theorem, $\sqrt{n}(\widehat{\theta}-\theta)$ is asymptotically multivariate Gaussian as $n \to \infty$, with mean zero and covariance matrix $E(B^{-1}MB^{-1})$, as defined in equation (4)(5)

For FSA 4 (Mancl & DeRouen estimator),

[[This is the version of proof for prototypical cSMART]] In the robust sandwich estimator in cluster GEE without the counterfactual summary term in cSMART, $r_i = \widehat{\epsilon}_i(X_i,Y_i;\hat{\beta},\hat{\eta}) = Y_i - \mu(X_i;\hat{\beta},\hat{\eta})$ is used to estimate $cov(Y_i)$ by $cov(Y_i) = r_i r_i^T$.

Here in cSMART, when calculating $M = E(U_{i,\beta,\eta}U_{i,\beta,\eta}^T)$, if the cluster is consistent with multiple AIs, then the middle term include cross multiplication across AIs, and the corresponding terms are in the form of $r_{i,(a_1,a_2)}r_{i,(a_1',a_2')}^T$ ($(a_1,a_2)$ and $(a_1',a_2')$ can be identical or not). Where $r_{i,(a_1,a_2)} = \widehat{\epsilon}_{i,(a_1,a_2)}(X_i,Y_i;\hat{\beta},\hat{\eta}) = Y_i - \mu(a_1,a_2,X_i;\hat{\beta},\hat{\eta})$.

However, it is argued that this estimation is often closer to the zero than the true value $\epsilon_{i,(a_1,a_2)}$, so we consider a first-order Taylor expansion $r_{i,(a_1,a_2)} = \epsilon_{i,(a_1,a_2)} + \frac{\partial \epsilon_{i,(a_1,a_2)}}{\partial (\beta,\eta)^T}((\hat{\beta},\hat{\eta})-(\beta,\eta))$, where $\epsilon_{i,(a_1,a_2)} = Y_i - \mu(a_1,a_2,X_i;\beta,\eta)$

From previous proof, we also have the first-order approximation

$$\widehat{\theta}-\theta = \tilde{B}^{-1}(\frac{1}{n}\sum_{i=1}^{n}\sum_{(a_1,a_2)} I_{i,a_1,a_2}(S_i)W_{i,a_1,a_2}(X_i,S_i)D_{a_1,a_2}(X_i)^T V_{i,a_1,a_2}^{-1}(X_i,\alpha)(Y_i-\mu(a_1,a_2,X_i;\theta)))+$$

$o(1)$

We define $cov(Y_i)_{(a_1,a_2),(a_1',a_2')} = \epsilon_{i,(a_1,a_2)}\epsilon_{i,(a_1',a_2')}^T$

Then $E[r_{i,(a_1,a_2)}r_{i,(a_1',a_2')}^T] = E[\epsilon_{i,(a_1,a_2)}\epsilon_{i,(a_1',a_2')}^T] + E[\epsilon_{i,(a_1,a_2)}(\widehat{\theta}-\theta)^T\frac{\partial \epsilon_{i,(a_1',a_2')}^T}{\partial \theta}]+$
$E[\frac{\partial \epsilon_{i,(a_1,a_2)}}{\partial \theta^T}(\widehat{\theta}-\theta)\epsilon_{i,(a_1',a_2')}^T] + E[\frac{\partial \epsilon_{i,(a_1,a_2)}}{\partial \theta^T}(\widehat{\theta}-\theta)(\widehat{\theta}-\theta)^T\frac{\partial \epsilon_{i,(a_1',a_2')}^T}{\partial \theta}]$

We define $H_{i,j,a_1,a_2,a_1',a_2'} = D_{a_1,a_2}(X_i)\,B^{-1}\,D_{a_1',a_2'}(X_j)^T V_{j,a_1',a_2'}^{-1}$

By neglecting the terms in the form $H_{i,j,a_1,a_2,a_1',a_2'}$ where $i \neq j$ (which is justified by simulation that such between-cluster terms are neglectable), the remaining terms are:

$$\asymp cov(Y_i)_{(a_1,a_2),(a'_1,a'_2)}-$$

$$cov(Y_i)_{(a_1,a_2),(a_1,a_2)}H^T_{i,i,a'_1,a'_2,a_1,a_2}W_{i,a_1,a_2}(X_i,S_i)-cov(Y_i)_{(a_1,a_2),(a'_1,a'_2)}H^T_{i,i,a'_1,a'_2,a'_1,a'_2}W_{i,a'_1,a'_2}(X_i,S_i)-$$

$$H_{i,i,a_1,a_2,a_1,a_2}cov(Y_i)_{(a_1,a_2),(a'_1,a'_2)}W_{i,a_1,a_2}(X_i,S_i)-H_{i,i,a_1,a_2,a'_1,a'_2}cov(Y_i)_{(a'_1,a'_2),(a'_1,a'_2)}W_{i,a'_1,a'_2}(X_i,S_i)+$$

$$H_{i,i,a_1,a_2,a_1,a_2}cov(Y_i)_{(a_1,a_2),(a_1,a_2)}H^T_{i,i,a'_1,a'_2,a_1,a_2}W_{i,a_1,a_2}(X_i,S_i)W_{i,a_1,a_2}(X_i,S_i)+$$

$$H_{i,i,a_1,a_2,a_1,a_2}cov(Y_i)_{(a_1,a_2),(a'_1,a'_2)}H^T_{i,i,a'_1,a'_2,a'_1,a'_2}W_{i,a_1,a_2}(X_i,S_i)W_{i,a'_1,a'_2}(X_i,S_i)+$$

$$H_{i,i,a_1,a_2,a'_1,a'_2}cov(Y_i)_{(a'_1,a'_2),(a_1,a_2)}H^T_{i,i,a'_1,a'_2,a_1,a_2}W_{i,a'_1,a'_2}(X_i,S_i)W_{i,a_1,a_2}(X_i,S_i)+$$

$$H_{i,i,a_1,a_2,a'_1,a'_2}cov(Y_i)_{(a'_1,a'_2),(a'_1,a'_2)}H^T_{i,i,a'_1,a'_2,a'_1,a'_2}W_{i,a'_1,a'_2}(X_i,S_i)W_{i,a'_1,a'_2}(X_i,S_i)$$

We do similar expansion for $E[r_{i,(a_1,a_2)}r^T_{i,(a_1,a_2)}]$, $E[r_{i,(a'_1,a'_2)}r^T_{i,(a_1,a_2)}]$ and $E[r_{i,(a'_1,a'_2)}r^T_{i,(a'_1,a'_2)}]$ and calculate the weighted sum of the corresponding equations together, and get the following formula:

$$\sum_{(a_1,a_2)}\sum_{(a'_1,a'_2)}I_{i,a_1,a_2}(S_i)I_{i,a'_1,a'_2}(S_i)W_{i,a_1,a_2}(X_i,S_i)W_{i,a'_1,a'_2}(X_i,S_i)r_{i,(a_1,a_2)}r^T_{i,(a'_1,a'_2)}$$
$$\asymp \sum_{(a_1,a_2)}\sum_{(a'_1,a'_2)}I_{i,a_1,a_2}(S_i)I_{i,a'_1,a'_2}(S_i)W_{i,a_1,a_2}(X_i,S_i)W_{i,a'_1,a'_2}(X_i,S_i)$$
$$(I_{m_i}-\sum_{(b_1,b_2)}I_{i,b_1,b_2}(S_i)W_{i,b_1,b_2}(X_i,S_i)H_{i,i,b_1,b_2,a_1,a_2})\epsilon_{i,(a_1,a_2)}\epsilon^T_{i,(a'_1,a'_2)}$$
$$(I_{m_i}-\sum_{(b'_1,b'_2)}I_{i,b'_1,b'_2}(S_i)W_{i,b'_1,b'_2}(X_i,S_i)H_{i,i,b'_1,b'_2,a'_1,a'_2})^T$$

So we have the approximation:

$$\sum_{(a_1,a_2)}\sum_{(a'_1,a'_2)}I_{i,a_1,a_2}(S_i)I_{i,a'_1,a'_2}(S_i))W_{i,a_1,a_2}(X_i,S_i)W_{i,a'_1,a'_2}(X_i,S_i)\epsilon_{i,(a_1,a_2)}\epsilon^T_{i,(a'_1,a'_2)} \asymp$$
$$\sum_{(a_1,a_2)}\sum_{(a'_1,a'_2)}I_{i,a_1,a_2}(S_i)I_{i,a'_1,a'_2}(S_i))W_{i,a_1,a_2}(X_i,S_i)W_{i,a'_1,a'_2}(X_i,S_i)$$
$$(I_{m_i}-\sum_{(b_1,b_2)}I_{i,b_1,b_2}(S_i)W_{i,b_1,b_2}(X_i,S_i)H_{i,i,b_1,b_2,a_1,a_2})^{-1}r_{i,(a_1,a_2)}r^T_{i,(a'_1,a'_2)}$$
$$(I_{m_i}-\sum_{(b'_1,b'_2)}I_{i,b'_1,b'_2}(S_i)W_{i,b'_1,b'_2}(X_i,S_i)H_{i,i,b'_1,b'_2,a'_1,a'_2})^{-T}$$

Finally, we have the following approximation:

$$(\sum_{i=1}^n\sum_{(a_1,a_2)}I_{i,a_1,a_2}(S_i)W_{i,a_1,a_2}(X_i,S_i)D_{a_1,a_2}(X_i)^TV^{-1}_{i,a_1,a_2}(X_i,\alpha)\epsilon_{i,(a_1,a_2)})^{\otimes 2}=$$
$$(\sum_{i=1}^n\sum_{(a_1,a_2)}I_{i,a_1,a_2}(S_i)W_{i,a_1,a_2}(X_i,S_i)D_{a_1,a_2}(X_i)^TV^{-1}_{i,a_1,a_2}(X_i,\alpha)$$
$$(I_{m_i}-\sum_{(b_1,b_2)}I_{i,b_1,b_2}(S_i)W_{i,b_1,b_2}(X_i,S_i)H_{i,i,b_1,b_2,a_1,a_2})^{-1}r_{i,(a_1,a_2)})^{\otimes 2}$$

If one cluster is consistent with more than two AIs, similar summation still apply and give identical results.

———————————

[[this version is about adjustment directly derived to use in general setting (any structure of SMART)]] To generalize the proof to more flexible setting, we use notation $a=1,2,...,A$ as indicator of the embedded Adaptive Intervention (substitute for

$(a_1, a_2)$ in prototypical SMART. We use $\theta = \theta$ to denote the combination of parameters in the marginal mean model.

In the robust sandwich estimator in cluster GEE without the counterfactual summary term in cSMART, $r_i = \widehat{\epsilon}_i(X_i, Y_i; \hat{\theta}) = Y_i - \mu(X_i; \hat{\theta})$ is used to estimate $cov(Y_i)$ by $cov(Y_i) = r_i r_i^T$.

Here in cSMART, when calculating $M = E(U_{i,\theta} U_{i,\theta}^T)$, if the cluster is consistent with multiple AIs, then the middle term include cross multiplication across AIs, and the corresponding terms are in the form of $r_{i,a} r_{i,a'}^T$ ($a$ and $a'$ can be identical or not). Where $r_{i,a} = \widehat{\epsilon}_{i,a}(X_i, Y_i; \hat{\theta}) = Y_i - \mu(a, X_i; \hat{\theta})$.

However, it is argued that this estimation is often closer to the zero than the true value $\epsilon_{i,a}$, so we consider a first-order Taylor expansion $r_{i,a} = \epsilon_{i,a} + \frac{\partial \epsilon_{i,a}}{\partial \theta^T}(\hat{\theta} - \theta)$, where $\epsilon_{i,a} = Y_i - \mu(a, X_i; \theta)$

Also from previous proof, we also have the first-order approximation
$$\hat{\theta} - \theta = -(n\tilde{B})^{-1}(\sum_{i=1}^{n} \sum_{a} I_{i,a}(S_i) W_{i,a}(X_i, S_i) D_a(X_i)^T V_{i,a}^{-1}(X_i, \alpha)(Y_i - \mu(a, X_i; \theta))) + o(1)$$

We define $cov(Y_i)_{a,a'} = \epsilon_{i,a} \epsilon_{i,a'}^T$

Then $E[r_{i,a} r_{i,a'}^T] = E[\epsilon_{i,a} \epsilon_{i,a'}^T] + E[\epsilon_{i,a}(\hat{\theta} - \theta)^T \frac{\partial \epsilon_{i,a'}^T}{\partial \theta}] +$
$E[\frac{\partial \epsilon_{i,a}}{\partial \theta^T}(\hat{\theta} - \theta) \epsilon_{i,a'}^T] + E[\frac{\partial \epsilon_{i,a}}{\partial \theta^T}(\hat{\theta} - \theta)(\hat{\theta} - \theta)^T \frac{\partial \epsilon_{i,a'}^T}{\partial \theta}]$

We define $H_{i,j,a,a'} = D_a(X_i)(nB)^{-1} D_{a'}(X_j)^T V_{j,a'}^{-1}$

By neglecting the terms in the form $H_{i,j,a,a'}$ where $i \neq j$ (which is justified by simulation that such between-cluster terms are neglectable), the remaining terms are:

$\asymp cov(Y_i)_{a,a'} -$
$\sum_{b} I_{i,b}(S_i) W_{i,b}(X_i, S_i) cov(Y_i)_{a,b} H_{i,i,a',b}^T -$
$\sum_{b} I_{i,b}(S_i) W_{i,b}(X_i, S_i) H_{i,i,a,b} cov(Y_i)_{b,a'} +$
$\sum_{b} \sum_{b'} I_{i,b}(S_i) I_{i,b'}(S_i) W_{i,b}(X_i, S_i) W_{i,b'}(X_i, S_i) H_{i,i,a,b} cov(Y_i)_{b,b'} H_{i,i,a',b'}^T$

We do similar expansion for all $E[r_{i,a} r_{i,a'}^T]$, insert modification matrix $Q_{i,a}$ and calculate the weighted sum of the corresponding equations together, and get the following formula:

$\sum_{a} \sum_{a'} I_{i,a}(S_i) I_{i,a'}(S_i) W_{i,a}(X_i, S_i) W_{i,a'}(X_i, S_i) Q_{i,a} r_{i,a} r_{i,a'}^T Q_{i,a'}^T$
$\asymp \sum_{a} \sum_{a'} I_{i,a}(S_i) I_{i,a'}(S_i) W_{i,a}(X_i, S_i) W_{i,a'}(X_i, S_i)$
$(Q_{i,a} - \sum_{b} I_{i,b}(S_i) W_{i,b}(X_i, S_i) Q_{i,b} H_{i,i,b,a}) \epsilon_{i,a} \epsilon_{i,a'}^T$

$$(Q_{i,a'} - \sum_{b'} I_{i,b'}(S_i)W_{i,b'}(X_i,S_i)Q_{i,b'}H_{i,i,b',a'})^T$$

We expect such summation to be equal to:

$$\sum_a \sum_{a'} I_{i,a}(S_i)I_{i,a'}(S_i)W_{i,a}(X_i,S_i)W_{i,a'}(X_i,S_i)$$
$$D_a(X_i)^T V_{i,a}^{-1}(X_i,\alpha)\epsilon_{i,a}\epsilon_{i,a'}^T V_{i,a'}^{-1} D_{a'}(X_i)$$

A sufficient condition is:

$$\Leftarrow Q_{i,a} - \sum_b I_{i,b}(S_i)W_{i,b}(X_i,S_i)Q_{i,b}H_{i,i,b,a} = D_a(X_i)^T V_{i,a}^{-1}(X_i,\alpha)$$
$$\Leftrightarrow Q_{i,a} - \sum_b I_{i,b}(S_i)W_{i,b}(X_i,S_i)Q_{i,b}D_b(X_i)(nB)^{-1}D_a(X_i)^T V_{i,a}^{-1}(X_i,\alpha) = D_a(X_i)^T V_{i,a}^{-1}(X_i,\alpha)$$

We set $Q_{i,a} = \tilde{Q}_{i,a} D_a(X_i)^T V_{i,a}^{-1}(X_i,\alpha)$, then

$$\Leftarrow \tilde{Q}_{i,a} = \sum_b I_{i,b}(S_i)W_{i,b}(X_i,S_i)\tilde{Q}_{i,b}D_b(X_i)^T V_{i,b}^{-1}(X_i,\alpha)D_b(X_i)(nB)^{-1} + I_{q+p}$$

This indicates $\tilde{Q}_{i,a}$ is independent of $a$.

So we have the solution:

$$\tilde{Q}_{i,a} = (I_{q+p} - \sum_b I_{i,b}(S_i)W_{i,b}(X_i,S_i)D_b(X_i)^T V_{i,b}^{-1}(X_i,\alpha)D_b(X_i)(nB)^{-1})^{-1}$$
$$Q_{i,a} = (I_{q+p} - \sum_b I_{i,b}(S_i)W_{i,b}(X_i,S_i)D_b(X_i)^T V_{i,b}^{-1}(X_i,\alpha)D_b(X_i)(nB)^{-1})^{-1} D_a(X_i)^T V_{i,a}^{-1}(X_i,\alpha)$$

Plugging back into the variance estimator, we get our version of FSA 4.

Our methods is an extension to the McCaffrey's in a way that the direct extension of their method in SMART setting is trying to ensure $E((\sum_a I_{i,a} W_{i,a}(X_i,S_i) Q_{i,a} r_{i,a})^{\otimes 2})$ is close to $E((\sum_a I_{i,a} W_{i,a}(X_i,S_i)\epsilon_{i,a})^{\otimes 2})$, while we take further consideration of the components of GEE equation (i.e. $D_a(X_i)^T V_{i,a}^{-1}(X_i)$) and want the adjusted outer product to be close to $E((\sum_a I_{i,a} W_{i,a}(X_i,S_i) D_a(X_i)^T V_{i,a}^{-1}(X_i) \epsilon_{i,a})^{\otimes 2})$.

## A.3 Appendix 3. Proof that the variance estimator remains the same when we don't consider cluster-level covariates

When we don't consider cluster-level covariates in our working model, we notice that the two sandwich estimators we propose remain the same no matter we assume independence or marginal exchangeable. As these two assumptions can be viewed as $\rho = 0$ vs $\rho \in [0,1]$, we only need to prove that the sandwich estimators do not contain $\rho$. The proof of such a phenomena is below.

We showed in Appendix 1 that the nonsingular linear transformation of $\beta_0, \ldots, \beta_3$ does not affect the statistical properties of interest, so we now use $\mu_{a_1,a_2}$ as the parameters in the mean model. Therefore, we expect the variance estimator to be zeros in its (1,3),(1,4),(2,3),(2,4) entries, as three treatment pathways merge into AI(1,$a_2$), and have no intersection with the other three pathways that merge into AI(-1,$a_2$).

So we now focus only on AI(1,$a_2$), and assume there are $n_l$ clusters in the $l_{th}$ treatment sequence (l=1,2,3). We denote $l_i$ as the sequence that cluster i belongs to.

(I) We prove that $\mu_{a_1,a_2}$ and $\sigma^2_{a_1,a_2}$ remain the same.

The GEE equation in this setting becomes:
$$2\sum_{l_i=1}\begin{pmatrix}\mathbf{1}_m & \mathbf{0}_m & \mathbf{0}_m & \mathbf{0}_m\end{pmatrix}^T \frac{1}{\sigma^2_{1,1}}CS_m(\rho_{1,1})^{-1}(Y_i - \mu_{1,1}\mathbf{1}_m)+$$
$$4\sum_{l_i=2}\begin{pmatrix}\mathbf{1}_m & \mathbf{0}_m & \mathbf{0}_m & \mathbf{0}_m\end{pmatrix}^T \frac{1}{\sigma^2_{1,1}}CS_m(\rho_{1,1})^{-1}(Y_i - \mu_{1,1}\mathbf{1}_m)+$$
$$2\sum_{l_i=1}\begin{pmatrix}\mathbf{0}_m & \mathbf{1}_m & \mathbf{0}_m & \mathbf{0}_m\end{pmatrix}^T \frac{1}{\sigma^2_{1,-1}}CS_m(\rho_{1,-1})^{-1}(Y_i - \mu_{1,-1}\mathbf{1}_m)+$$
$$4\sum_{l_i=2}\begin{pmatrix}\mathbf{0}_m & \mathbf{1}_m & \mathbf{0}_m & \mathbf{0}_m\end{pmatrix}^T \frac{1}{\sigma^2_{1,-1}}CS_m(\rho_{1,-1})^{-1}(Y_i - \mu_{1,-1}\mathbf{1}_m) = 0$$

Directly looking into the equation we can see $\mu_{1,1} = \frac{\sum_{i=1}^n I_{i,1,1}W_i \sum_{j=1}^m Y_{ij}}{\sum_{i=1}^n I_{i,1,1}W_i}$ is the solution, thus the only solution, and it is independent of $\rho$.

Therefore, $\widehat{\epsilon}_{ij}(a_1, a_2|) = Y_{ij} - \mu_{a_1,a_2}$ is independent of $\rho$, and $\widehat{\sigma}^2_{a1,a2} = \frac{\sum_{i=1}^n I_{i,a_1,a_2}W_i \sum_{j=1}^m \widehat{\epsilon}^2_{ij}(a_1,a_2|)}{\sum_{i=1}^n I_{i,a_1,a_2}W_i m}$ is independent of $\rho$.

(II) We prove that $V_{LZ}$ are the same.
$$\widehat{V}_{LZ} = \widehat{V}_{naive}(\sum_{i=1}^n U_i(A_{1i}, R_i, A_{2i}, X_i, Y_i; \beta, \eta|U_i^T(A_{1i}, R_i, A_{2i}, X_i, Y_i; \beta, \eta|)\widehat{V}_{naive}$$
$$= \widehat{V}_{naive}\sum_{i=1}^n(\sum_{(a_1,a_2)} I_{i,a_1,a_2}W_i D(X_i, a_1, a_2|^T \widehat{V}^{-1}_{i,a_1,a_2}(Y_i - \mu(x_i, a_1, a_2; \beta, \eta|))^{\otimes 2}\widehat{V}_{naive}$$
First, Let's see what the inner part is.

if $l_i = 2$,
$$U_i = 4\begin{pmatrix}\mathbf{1}_m & \mathbf{0}_m & \mathbf{0}_m & \mathbf{0}_m\end{pmatrix}^T \frac{1}{\sigma^2_{1,1}}CS_m(\rho_{1,1})^{-1}(Y_i - \mu_{1,1}\mathbf{1}_m)$$

$$= \frac{4}{\sigma_{1,1}^2} \begin{pmatrix} \mathbf{1}_m^T \\ \mathbf{0}_m^T \\ \mathbf{0}_m^T \\ \mathbf{0}_m^T \end{pmatrix} (\frac{1}{1-\rho_{1,1}} I_m - \frac{\rho_{1,1}}{(1-\rho_{1,1})(1+(m-1)\rho_{1,1})} J_m)(Y_i - \mu_{1,1}\mathbf{1}_m)$$

$$= \frac{4}{\sigma_{1,1}^2} \left( (\frac{1}{1-\rho_{1,1}}\mathbf{1}_m - \frac{m\rho_{1,1}}{(1-\rho_{1,1})(1+(m-1)\rho_{1,1})}\mathbf{1}_m), \ \mathbf{0}_m, \ \mathbf{0}_m, \ \mathbf{0}_m \right)^T (Y_i - \mu_{1,1}\mathbf{1}_m)$$

$$= \frac{4}{\sigma_{1,1}^2(1+(m-1)\rho_{1,1})} \begin{pmatrix} \mathbf{1}_m & \mathbf{0}_m & \mathbf{0}_m & \mathbf{0}_m \end{pmatrix}^T (Y_i - \mu_{1,1}\mathbf{1}_m) = \frac{4}{\sigma_{1,1}^2(1+(m-1)\rho_{1,1})} \left( \sum_{j=1}^m Y_{ij} - m\mu_{1,1}, \ 0, \ 0, \ 0 \right)^T$$

$$U_i^{\otimes 2} = \frac{16m^2}{\sigma_{1,1}^4(1+(m-1)\rho_{1,1})^2} \begin{pmatrix} (\frac{1}{m}\sum_{j=1}^m Y_{ij} - \mu_{1,1})^2 & 0 & 0 & 0 \\ 0 & 0 & 0 & 0 \\ 0 & 0 & 0 & 0 \\ 0 & 0 & 0 & 0 \end{pmatrix}$$

if $l_i = 1$,

$$U_i = 2\begin{pmatrix} \mathbf{1}_m & \mathbf{0}_m & \mathbf{0}_m & \mathbf{0}_m \end{pmatrix}^T \frac{1}{\sigma_{1,1}^2} CS_m(\rho_{1,1})^{-1}(Y_i - \mu_{1,1}\mathbf{1}_m)$$
$$+ 2\begin{pmatrix} \mathbf{0}_m & \mathbf{1}_m & \mathbf{0}_m & \mathbf{0}_m \end{pmatrix}^T \frac{1}{\sigma_{1,-1}^2} CS_m(\rho_{1,-1})^{-1}(Y_i - \mu_{1,-1}\mathbf{1}_m)$$
$$= \left( \frac{2m}{\sigma_{1,1}^2(1+(m-1)\rho_{1,1})}(\frac{1}{m}\sum_{j=1}^m Y_{ij} - \mu_{1,1}), \ \frac{2m}{\sigma_{1,-1}^2(1+(m-1)\rho_{1,-1})}(\frac{1}{m}\sum_{j=1}^m Y_{ij} - \mu_{1,-1}), \ 0, \ 0 \right)^T$$

$$U_i^{\otimes 2} = \begin{pmatrix} \frac{4m^2}{\sigma_{1,1}^4(1+(m-1)\rho_{1,1})^2}(\frac{1}{m}\sum_{j=1}^m Y_{ij} - \mu_{1,1})^2 & U_{12} & 0 & 0 \\ U_{12} & \frac{4m^2}{\sigma_{1,-1}^4(1+(m-1)\rho_{1,-1})^2}(\frac{1}{m}\sum_{j=1}^m Y_{ij} - \mu_{1,-1})^2 & 0 & 0 \\ 0 & 0 & 0 & 0 \\ 0 & 0 & 0 & 0 \end{pmatrix}$$

Here $U_{12} = \frac{4m^2}{\sigma_{1,1}^2\sigma_{1,-1}^2(1+(m-1)\rho_{1,1})(1+(m-1)\rho_{1,-1})}(\frac{1}{m}\sum_{j=1}^m Y_{ij} - \mu_{1,1})(\frac{1}{m}\sum_{j=1}^m Y_{ij} - \mu_{1,-1})$

Therefore, the diagonal term of the meat of the sandwich estimator is:

$$\frac{m^2}{\sigma_{a_1,a_2}^4(1+(m-1)\rho_{a_1,a_2})^2} \sum_{i=1}^n I_{i,a_1,a_2} W_i^2 (\frac{1}{m}\sum_{j=1}^m Y_{ij} - \mu_{a_1,a_2})^2$$

and the (1,2) term and (3,4) term is:

$$\frac{m^2}{\sigma_{a_1,1}^2\sigma_{a_1,-1}^2(1+(m-1)\rho_{a_1,1})(1+(m-1)\rho_{a_1,-1})} \sum_{i=1}^n I_{i,a_1,1} I_{i,a_1,-1} W_i^2 (\frac{1}{m}\sum_{j=1}^m Y_{ij} - \mu_{a_1,1})(\frac{1}{m}\sum_{j=1}^m Y_{ij} - \mu_{a_1,-1})$$

For $\widehat{V}_{naive} = (\sum_{i=1}^n \sum_{(a_1,a_2)} I_{i,a_1,a_2} W_i D(X_i, a_1, a_2|^T \widehat{V}_{i,a_1,a_2}^{-1} D(X_i, a_1, a_2|))^{-1}$, similarly we have:

if $l_i = 2$,

$$\sum_{(a_1,a_2)} I_{i,a_1,a_2} W_i D(X_i, a_1, a_2|^T \widehat{V}_{i,a_1,a_2}^{-1} D(X_i, a_1, a_2| = W_i D(X_i, 1, 1|^T \widehat{V}_{i,1,1}^{-1} D(X_i, 1, 1|$$

$$= \begin{pmatrix} \mathbf{1}_m & \mathbf{0}_m & \mathbf{0}_m & \mathbf{0}_m \end{pmatrix}^T \frac{1}{\sigma_{1,1}^2} CS_m(\rho_{1,1})^{-1} \begin{pmatrix} \mathbf{1}_m & \mathbf{0}_m & \mathbf{0}_m & \mathbf{0}_m \end{pmatrix} = \frac{1}{\sigma_{1,1}^2(1+(m-1)\rho_{1,1})} \begin{pmatrix} 1 & 0 & 0 & 0 \\ 0 & 0 & 0 & 0 \\ 0 & 0 & 0 & 0 \\ 0 & 0 & 0 & 0 \end{pmatrix}$$

Because there is no outer product of vector, thus the formula is additive, so we can conclude that $\widehat{V}_{naive}$ is diagonal, and the diagonal term is $(\frac{1}{\sigma_{a_1,a_2}^2(1+(m-1)\rho_{a_1,a_2})} \sum_{i=1}^{n} I_{i,a_1,a_2} W_i)^{-1}$

Therefore, in $\widehat{V}_{LZ}$, all terms concerning $\sigma_{a_1,a_2}^2$ and $(1+(m-1)\rho_{a_1,a_2})$ cancel off, and the final formula does not contain $\rho_{a_1,a_2}$.

(III) We prove that $V_{MD}$ are the same

$$\widehat{V}_{MD} = \widehat{V}_{naive} \sum_{i=1}^{n} (\sum_{(a_1,a_2)} I_{i,a_1,a_2} W_i D(X_i, a_1, a_2|^T \widehat{V}_{i,a_1,a_2}^{-1} (I_{m_i} - H_{i,a_1,a_2})^{-1} (Y_i - \mu(x_i, a_1, a_2; \beta, \eta||))^{\otimes 2} \widehat{V}_{naive}$$

where $H_{i,a_1,a_2} = D(X_i, a_1, a_2| \widehat{V}_{naive} D(X_i, a_1, a_2|^T \widehat{V}_{i,a_1,a_2}^{-1}$

Take $(a_1, a_2) = (1, 1)$ as example,

$$H_{i,1,1} = \begin{pmatrix} \mathbf{1}_m & \mathbf{0}_m & \mathbf{0}_m & \mathbf{0}_m \end{pmatrix} V_{naive} \begin{pmatrix} \mathbf{1}_m & \mathbf{0}_m & \mathbf{0}_m & \mathbf{0}_m \end{pmatrix}^T \frac{1}{\sigma_{1,1}^2} CS_m(\rho_{1,1})^{-1}$$

$$= (V_{naive})_{(1,1)} \cdot \frac{1}{\sigma_{1,1}^2(1+(m-1)\rho_{1,1})} J_m = \frac{1}{\sum_{i=1}^{n} I_{i,1,1} W_i} J_m$$

We denote $\omega_{a_1,a_2} \triangleq \sum_{i=1}^{n} I_{i,a_1,a_2} W_i$, then $H_{i,a_1,a_2} = \frac{1}{\omega_{a_1,a_2}} J_m$

$$(I_m - H_{i,a_1,a_2})^{-1} = (I_m - \frac{1}{\omega_{a_1,a_2}} J_m)^{-1} = I_m + \frac{1}{\omega_{a_1,a_2}-m} J_m$$

Similarly, if $l_i = 2$,

$$U_i = \frac{4}{\sigma_{1,1}^2(1+(m-1)\rho_{1,1})} \begin{pmatrix} \mathbf{1}_m & \mathbf{0}_m & \mathbf{0}_m & \mathbf{0}_m \end{pmatrix}^T (I_m + \frac{1}{\omega_{1,1}-m} J_m)(Y_i - \mu_{1,1} \mathbf{1}_m)$$

$$= \frac{4}{\sigma_{1,1}^2(1+(m-1)\rho_{1,1})} \frac{\omega_{1,1}}{\omega_{1,1}-m} \begin{pmatrix} \sum_{j=1}^{m} Y_{ij} - m\mu_{1,1}, & 0, & 0, & 0 \end{pmatrix}^T$$

Comparing with the formula of $V_{LZ}$, we find $V_{MD} = W \cdot V_{LZ} \cdot W$, where W is a diagonal matrix with diagonal term $\frac{\omega_{a_1,a_2}}{\omega_{a_1,a_2}-m} = \frac{\sum_{i=1}^{n} I_{i,a_1,a_2} W_i}{\sum_{i=1}^{n} I_{i,a_1,a_2} W_i - m}$

Therefore, the formula of $V_{MD}$ does not contain $\rho_{a_1,a_2}$.

## A.4 Appendix 4. Derivation of formula to convert treatment-pathway level parameter to AI level

For notational issue, we use ①−⑥ to represent the 6 treatment pathway within a prototypical cSMART. We assume $Y_i \in \mathbb{R}^{m_i*1}$ as the observed outcome of cluster i, and $Y_i = \mu_l + \epsilon_i + \eta^T X_i * \mathbf{1}_{m_i}$, where $l \in 1, 2, ..., 6$ represent the treatment pathway where the cluster goes under. We use subscript $a_1, a_2$ to denote parameter at marginal (adaptive intervention) level, and subscript $l$ to denote parameter at treatment pathway level. For simplicity, we now consider how parameters in AI(1,1) is derived from parameters in treatment pathway ① and ②. We use p to denote response rate of corresponding first-stage treatment $a_1 = 1$. We use a single Y without subscript to denote a single observation (rather than the vector of result of a cluster).

For the marginal mean,
$$\mu_{1,1} = \mathop{\mathbb{E}}_{Y \sim AI(1,1)} Y = p * \mathop{\mathbb{E}}_{Y \sim ①} Y + (1-p) * \mathop{\mathbb{E}}_{Y \sim ②} Y = p\mu_1 + (1-p)\mu_2$$

For the marginal variance,
$$\sigma_{1,1}^2 = Var(Y_{1,1}) - Var(\eta^T X)$$
$$= \mathop{\mathbb{E}}_{Y \sim AI(1,1)} Y^2 - (\mathop{\mathbb{E}}_{Y \sim AI(1,1)} Y)^2 - Var(\eta^T X) = p * \mathop{\mathbb{E}}_{Y \sim ①} Y^2 + (1-p) * \mathop{\mathbb{E}}_{Y \sim ②} Y^2 - \mu_{1,1}^2 - Var(\eta^T X)$$
$$= p*(\mu_1^2 + \sigma_1^2 + Var(\eta^T X)) + (1-p)*(\mu_2^2 + \sigma_2^2 + Var(\eta^T X)) - (p\mu_1 + (1-p)\mu_2)^2 - Var(\eta^T X)$$
$$= p\sigma_1^2 + (1-p)\sigma_2^2 + p(1-p)(\mu_1 - \mu_2)^2$$

For the marginal ICC,

We further constrain Y and $Y'$ be in the same cluster, then
$$\mathop{\mathbb{E}}_{Y,Y' \sim AI(1,1)} YY' - \mathop{\mathbb{E}}_{Y \sim AI(1,1)} Y \mathop{\mathbb{E}}_{Y' \sim AI(1,1)} Y'$$
$$= p * \mathop{\mathbb{E}}_{Y,Y' \sim ①} YY' + (1-p) * \mathop{\mathbb{E}}_{Y,Y' \sim ②} YY' - p * \mu_1^2 - (1-p) * \mu_2^2$$
$$= p * (\mu_1^2 + \sigma_1^2 * \rho_1) + (1-p) * (\mu_2^2 + \sigma_2^2 * \rho_2) - p * \mu_1^2 - (1-p) * \mu_2^2$$
$$= p\sigma_1^2 * \rho_1 + (1-p)\sigma_2^2 * \rho_2$$

So $\rho_{1,1} = (p\sigma_1^2 * \rho_1 + (1-p)\sigma_2^2 * \rho_2)/(p\sigma_1^2 + (1-p)\sigma_2^2 + p(1-p)(\mu_1 - \mu_2)^2)$

## A.5 Appendix 5. Derivation of the modified variance estimator when estimating the inverse probability weights

Suppose $\hat{\gamma}$ is solved with estimating equation $\sum_{i=1}^{n} m_i(\gamma) = 0$, then

$$0 = \sum_{i=1}^{n} m_i(\hat{\gamma}) = \sum_{i=1}^{n} m_i(\gamma) + (\sum_{i=1}^{n} \frac{\partial m_i(\gamma)}{\partial \gamma}|_{\gamma=\hat{\gamma}})(\hat{\gamma} - \gamma)$$

Therefore, $\sqrt{n}(\hat{\gamma} - \gamma) = -\frac{1}{\sqrt{n}} (E\frac{\partial m(\gamma)}{\partial \gamma}|_{\gamma=\gamma})^{-1} \sum_{i=1}^{n} m_i(\gamma) + o_P(1)$

Let $S_{\gamma,i} = m_i(\gamma)$, by Information Equality, $E\frac{\partial m(\gamma)}{\partial \gamma}|_{\gamma=\gamma} = -E(m(\gamma)m(\gamma)^T)$

So $\sqrt{n}(\hat{\gamma} - \gamma) = \frac{1}{\sqrt{n}} \sum_{i=1}^{n} F^{-1} S_{\gamma,i} + o_P(1)$, where $F = E(S_{\gamma,i} S_{\gamma,i}^T)$

Therefore, $0 = \sum_{i=1}^{n} U_{i,\hat{\beta},\hat{\eta},\hat{\gamma}} = \sum_{i=1}^{n} U_{i,\beta,\eta,\gamma} + (\sum_{i=1}^{n} \frac{\partial U_i}{\partial (\beta,\eta)})|_{(\beta,\eta)=(\tilde{\beta},\tilde{\eta})}((\hat{\beta},\hat{\eta}) - (\beta,\eta)) + (\sum_{i=1}^{n} \frac{\partial U_i}{\partial \gamma})|_{\gamma=\tilde{\gamma}}(\hat{\gamma} - \gamma) + o_P(1)$

So $\sqrt{n}((\hat{\beta},\hat{\eta}) - (\beta,\eta)) = -(E\frac{\partial U}{\partial (\beta,\eta)})^{-1}(\frac{1}{\sqrt{n}} \sum_{i=1}^{n} U_{i,\beta,\eta,\gamma} + E(\frac{\partial U_{i,\beta,\eta}}{\partial \gamma})\sqrt{n}(\hat{\gamma} - \gamma)) + o_P(1)$

$= B^{-1}(\frac{1}{\sqrt{n}} \sum_{i=1}^{n} U_{i,\beta,\eta,\gamma} + C\frac{1}{\sqrt{n}} \sum_{i=1}^{n} F^{-1} S_{\gamma,i}) + o_P(1) = \frac{1}{\sqrt{n}} B^{-1} \sum_{i=1}^{n} (U_{i,\beta,\eta,\gamma} + CF^{-1}S_{\gamma,i}) + o_P(1)$

By Central Limit Theorem, $\sqrt{n}((\hat{\beta},\hat{\eta})-(\beta,\eta)) \xrightarrow{d} N(0, B^{-1}E((U_{i,\beta,\eta,\gamma}+CF^{-1}S_{\gamma,i})(U_{i,\beta,\eta,\gamma}+CF^{-1}S_{\gamma,i})^T)B^{-1})$

By no shared parameter of the two estimating equations, $\sqrt{n}((\hat{\beta},\hat{\eta}) - (\beta,\eta)) \xrightarrow{d} N(0, B^{-1}(M - CF^{-1}C)B^{-1})$

## A.6 Appendix 6. Results of utilizing modeling assumptions that are known to improve asymptotic efficiency

The purpose of this second set of simulation experiments was to understand whether the results differed when we contrast strategies that are known to improve statistical efficiency.

These known strategies include: (a) adjusting for baseline covariates (vs not adjusting); (b) using an exchangeable working variance assumption (vs an independent working assumption); (c) estimating the inverse probability weights (vs using known weights).

We evaluate then under varying parameter settings and summarize the results below:

With respect to (a) and (b), we find improved statistical efficiency, both in small and large samples, under data generative models in which (a) the baseline covariate being adjusted for is related to the outcome with effect sizes of approx. 0.2 or greater; and in which (b) the marginal ICC is approximately 0.03 or higher. We also find improved inference in these settings, regardless of sample size.

With respect to (c) using estimated vs known weights, the results depend on sample size. In small samples, there is efficiency loss due to estimated weights. In large samples, the results depend on the effect size: The larger effect size of main comparison between adaptive interventions, the smaller sample size needed to achieve efficiency gain.

For detailed numerical results, see supplementary material.